\documentstyle[12pt,twoside]{article}

\newtheorem{Th}{\hspace{\parindent}Theorem}

\newtheorem{Lem}{\hspace{\parindent}Lemma}

\newtheorem{Cor}{\hspace{\parindent}Corollary}[Th]

\newcommand{\rua}{Rois }
\newcommand{\binom}[2]     
{\left(
\raisebox{-0.9ex}{\mbox{$\stackrel {\textstyle {#1}}{#2}$}}
\right)}
\newcommand{\intt}[1]{\left\lfloor{#1}\right\rfloor} 
\newcommand{\sfrac}[2]{\frac{\textstyle {#1}}{#2}}
\newcommand{\mybar}[1]{\overline{#1}} 
\newcommand{\mybarr}[1]{\overline{#1}}
\newcommand{\mypi}{\mbox{\LARGE $\pi$}}  
\newcommand{\dsum}{\displaystyle\sum}    
\newcommand{\bc}{\begin{center}}
\newcommand{\ec}{\end{center}}

\begin{document}

\begin{titlepage}

\thispagestyle{empty}


\vspace*{70mm}

\bc
  \huge  \bf ON CRITICAL AND MAXIMAL \\ DIGRAPHS
\\ \ \\
\huge \bf G. \v{S}. Fridman
\ec

\vspace{50mm}

\bc
\large  Preprint N 99-1
\ec

\vspace{40mm}

\bc
\large   Omsk 1999
\ec

\end{titlepage}

\thispagestyle{empty}

\bc
\large \bf   Omsk State University
\ec

\vspace{50mm}

\bc
  \huge  \bf ON CRITICAL AND MAXIMAL \\ DIGRAPHS
\\ \ \\
\huge \bf G. \v{S}. Fridman
\ec

\vspace{50mm}

\bc
\large  Preprint N 99-1
\ec

\vspace{40mm}

\bc
\large   Omsk 1999
\ec

\newpage

\thispagestyle{empty}

\vspace*{30mm}

A reader may ask a natural question why this memoir has
appeared only now, more than twenty five years after publishing
a summary of basic results presented here [3].

The fact is that during those years I worked in A.A.Lyapunov's
Laboratory of Theore\-tical Cybernetics in Academic town in Novosibirsk.
The results published in [3] seemed to be very interesting for me
and I decided that their detailed proofs should be published in some
respectable international journal. But publishing of Soviet
mathematicians's works abroad was connected with certain formal
difficulties 25 years ago. Moreover, necessity of changing affiliation
after Lyapunov's death in 1973, leaving for Omsk in 1976, and new topics
of my research in the Omsk State University had put the
publishing this work far to the periphery. In 90-s I stopped
my work in mathematics and my activity focused on absolutely other
questions.

Probably this memoir would still remain unpublished but for the many
years interest of Professor Ferdinand Gliviak in these results.
For a long time he was studying similar problems and together with
a group of his disciples and colleagues got a number of
interesting results (see, for example, [5] -- [7]). Some time ago
he asked me again to publish the detailed proofs of results from [3].
I understood that I was indebted to my colleagues mathematicians.
And the debts are to be paid back.
At last
this text is in front of you.
But this would not happened, unless my friend and colleague
Dr. Victor Il'ev, who works actively in mathematics himself,
had found the power and time and had carried out the huge work on
preparation of this manuscript for publishing.

I am sincerely grateful for the mentioned above to Professor
Ferdinand Gliviak and Dr. Victor Il'ev, and also to Professor
Alexander Kolokolov for the opportunity to discuss different
scientific and other problems during 25 years.

\bigskip

\hspace*{100mm} {\large \it Author},

\hspace*{100mm} {\large \it January 1999}

\newpage

\begin{center}
\Large \bf  ON CRITICAL AND MAXIMAL DIGRAPHS
\\ \ \\
        G. \v{S}. Fridman
\footnote{E-mail: fridman.gennadiy@gmail.com}
\end{center}

\bigskip

\begin{abstract}

This paper is devoted to the study of directed graphs with extremal
properties
relative to certain metric functionals.
We  characterize up to isomorphism critical digraphs with infinite 
values of diameter, quasi-diameter, radius and quasi-radius.
Moreover, maximal
digraphs with finite values of radius and quasi-diameter
are studied.

\end{abstract}

\begin{center}
\large \bf
   Introduction
\end{center}

This paper is devoted to the study of directed graphs with extremal
properties
relative to
certain metric functionals.

We introduce a distance function $\rho(x,y)$ on the vertex set of a
directed graph without loops $G=(X,U)$ in the following way:
$\rho(x,x)=0$; for $x\ne y$ let $\rho(x,y)$ be equal to the minimum
number of arcs in a directed path from $x$ to $y$, if the vertex $y$
is a reachable from $x$; otherwise, we set $\rho(x,y)=\infty$.
We introduce in addition a
function $\rho_m(x,y)$ as follows:
$$
\rho_m(x,y)=\min\{\rho(x,y),\rho(y,x)\}.
$$
We define the following quantities of the digraph $G=(X,U)$:

\begin{tabbing}
\hspace{\parindent} \hspace{1cm}
\= the quasi-diameter gu \= $d_m(G)=\max\limits_{x,y\in X}\rho_m(x,y)$; \kill
\>the diameter \> $d(G)=\max\limits_{x,y\in X}\rho(x,y)$;\\
\>the quasi-diameter \> $d_m(G)=\max\limits_{x,y\in X}\rho_m(x,y)$;\\
\>the radius \> $r(G)= \min\limits_{x\in X}\max\limits_{y\in X}\rho(x,y)$;\\
\>the quasi-radius \> $r_m(G)= \min\limits_{x\in X}\max\limits_{y\in X}
\rho_m(x,y)$.
\end{tabbing}

A digraph $G=(X,U)$ is called $d$-{\em critical}, if the addition of an
arbitrary missing arc,
whose both endpoints
belong to $X$, results
either in decreasing the number of bicomponents (strong
components) or in decreasing the diameter. Analogously we can introduce
the notions of $d_m$-{\em critical}, $r$-{\em critical}, and
$r_m$-{\em critical} digraphs.

Let $i$ be an invariant of an $n$-vertex digraph $G$.
Then $G$ is called {\it maximal by} $i$, if it has the maximum number
of arcs among all $n$-vertex digraphs with invariant $i$.

In \cite{M70} $d$-critical digraphs were studied. Earlier, in \cite{O68},
the similar problem for undirected graphs was posed and solved.
In \cite{V67} the least upper bound is found
on the number of edges in an $n$-vertex ordinary undirected graph
with given finite radius and in \cite{I70}, \cite{I71}, upper bounds
on the number of arcs in digraphs of the following classes:
1) $n$-vertex digraphs without loops which have $k$ bicomponents
and infinite radius, 2) $n$-vertex digraphs without loops which have
$k$ bicomponents and given finite radius $r$ and are such that
each bicomponent is a complete symmetric digraph.

Graphs and digraphs, which are extremal with respect to deleting
of arcs and vertices, were studied in \cite{G75} -- \cite{GK95}.
Moreover, in \cite{GK95} some results on digraphs near to $r$-critical
digraphs were obtained, in \cite{GKS94} the similar problem for
undirected graphs was studied.

In \cite{F71} all $d$-critical digraphs of infinite diameter were
characterized up to isomorphism. In \cite{F73} $r$-critical digraphs
with infinite radius and $r_m$-critical digraphs with infinite
quasi-radius were characterized up to isomorphism. Moreover, in
\cite{F73} the least upper bound on the number of arcs in an $n$-vertex
digraph with given finite radius $r$ was obtained and
all maximal by $r$ digraphs were characterized up to isomorphism.
In \cite{F77} the similar results on maximal by $d_m$ digraphs
were presented.

This paper contains full proofs of all results presented in
\cite{F71} -- \cite{F77}.
In Sections 1, 2
we shall characterize up to
isomorphism critical digraphs with infinite values of $d, d_m, r$, and
$r_m$. In Section 3
maximal by $r$ and $d_m$ digraphs with finite values of $r$ and $d_m$
are studied.


\begin{center}
\large \bf
1.
 On critical digraphs with infinite values \\
  of diameter and quasi-diameter
\end{center}

In this section we shall characterize
up to isomorphism critical digraphs
with infinite values of diameter and quasi-diameter.

Two vertices $x$ and $y$ in a digraph $G=(X,U)$ are called {\em mutually
reachable\/}, if $\rho(x,y)<\infty$ and $\rho(y,x)<\infty$.
Obviously, the relation of mutual
reachability is an equivalence relation. Thus, the vertex set of $G$ is
partitioned into equivalence blocks
$$
X=X_1\cup...\cup X_k,\quad X_i \cap X_j=\emptyset \mbox{ for }i\not=j.
$$

The subgraph of the digraph $G$ induced by the set $X_i$ is called a
{\em strong component (bicomponent)\/}. A digraph that has
one bicomponent is called {\em biconnected\/}.

It is easy to see that
the diameter $d(G)$ of a digraph $G$ is finite if and only if
$G$ is biconnected. The quasi-diameter is finite if and only if for any
pair of vertices in $G$ at least one of them is reachable from the
other.

Let $G=(X,U)$ be a $d$-critical digraph and $d(G)=\infty$. Since the
operation of the transitive closure does not violate the relation of
mutual reachability, it is easy to see that $G$ is a transitive digraph.
Hence each bicomponent of $G$ is a complete symmetric digraph; and if an
arc goes in $G$ from some vertex of a bicomponent $A_i$ to some vertex of a
bicomponent $A_j$, then $G$ contains all arcs the initial vertices of which
belong to $A_i$ and the terminal ones belong to $A_j$.
The same is also valid for $d_m$-critical digraphs with infinite
quasi-diameter.

Let $D=(X,U)$ be an arbitrary digraph and $\{A_1,...,A_k\}$ be the
totality of its bicomponents. We associate with the digraph $D$ a
digraph $\Gamma(D)$ in the following way. The vertices of the digraph
$\Gamma(D)$ are $y_1,...,y_k$; an edge $(y_i,y_j)$ belongs to
$\Gamma(D)$ iff there is an arc in $D$ whose initial vertex belongs to
$A_i$ and the terminal vertex belongs to $A_j$. The digraph $\Gamma(D)$
is called the {\em Hertz graph\/} of the digraph $D$. Evidently,
$\Gamma(D)$ contains no directed cycles. Furthermore, if $D$ is a
transitive digraph, then its Hertz graph is also transitive.

The above considerations show that a $d$-critical digraph of
infinite diameter and a $d_m$-critical digraph of infinite
quasi-diameter can be uniquely up to isomorphism reconstructed from its
Hertz graph, whenever the number of vertices in each bicomponent is
determined. Hence the problem of describing $d$-critical digraphs
with $d=\infty$ and $d_m$-critical digraphs with $d_m=\infty$
is reduced to describing corresponding Hertz graphs. This is the topic of
the following considerations.

\begin{Lem} \label{lem1}
A transitive acyclic digraph is a subgraph of some complete transitive
acyclic digraph.
\end{Lem}

{\bf Proof.\/}
The set of arcs of a transitive acyclic digraph
defines on the
set of its vertices a relation of a partial order. Hence the assertion
of the lemma is equivalent to the well known fact that any partially
ordered finite set can be embedded into a linearly ordered set without
violating the partial order.
\rule{2mm}{2mm}

\bigskip

The totality of arcs of a complete acyclic transitive $k$-vertex digraph
defines on the set of its vertices a relation of linear order, therefore
it is isomorphic to the graph $\Gamma_k=(X_k,U_k)$ where
$X_k=\{1,...,k\}$ and $(i,j)\in U_k \Leftrightarrow i<j$. We shall call
a digraph that is isomorphic to the digraph $\Gamma_k$ for some $k$ a
{\em transitive tournament\/}.

We consider only directed graphs in this paper; henceforth the term
``graph'' will always mean directed graph.

\begin{Th} \label{th1}
The Hertz graph of a $d$-critical digraph of infinite diameter with
$k$ bicomponents is isomorphic to the graph $\Gamma_k$.
\end{Th}

{\bf Proof.\/}
Let a graph $\Gamma$ satisfy the hypothesis of the theorem; then it is
acyclic transitive digraph and hence, by Lemma \ref{lem1}, is isomorphic
to a subgraph $\Gamma'_k$ of the graph $\Gamma_k$. Let an arc $(i,j)$
where $i<j$ be missing in $\Gamma'_k$. Evidently the addition of the arc
does not result in the appearance of a directed cycle contrary to $\Gamma$
being $d$-critical with $d=\infty$. Hence $\Gamma'_k=\Gamma_k$.
\rule{2mm}{2mm}

\begin{Cor} \label{cor11}
Let $D$ be a nonbiconnected digraph with the property that the addition
of an arbitrary missing arc converts it into a biconnected digraph. Then
the Hertz graph of $D$ is isomorphic to the graph $\Gamma_2$.
\end{Cor}

{\bf Proof.\/}
By Theorem \ref{th1}, the Hertz graph of $D$ is isomorphic to the graph
$\Gamma_k$ for some $k$. But if $k\geq 3$, then the addition of the arc
$(k,k-1)$ to $\Gamma_k$ does not make it biconnected. Hence the Hertz
graph of the digraph $D$ is isomorphic to the graph $\Gamma_2$.
\rule{2mm}{2mm}

\bigskip

We note that the assertion of Corollary \ref{cor11} contains the \rua
theorem
\cite{Z69}.

\begin{Cor}[\rm See also \cite{I71}] \label{cor12}
The number of arcs in an $n$-vertex digraph of infinite diameter with
$k$ bicomponents does not exceed $n(n-k)+\sfrac{k^2-k}{2}$.
\end{Cor}

{\bf Proof.\/}
Let a digraph $G$ contain the greatest number of arcs among all
$n$-vertex digraphs of infinite diameter with $k$ bicomponents
$A_1,...,A_k$. Let the bicomponent $A_i$ contain $m_i$ vertices
$(i=1,...,k)$.

Evidently, $G$ is a $d$-critical digraph. By Theorem \ref{th1}, the
Hertz graph of $G$ is isomorphic to $\Gamma_k$. Let $2\leq m_i\leq m_j$.
Consider a $d$-critical digraph $G'$ of infinite diameter that is
obtained from $G$ by moving a vertex from the bicomponent $A_i$ to the
bicomponent $A_j$. In this case the number of arcs in $G'$ is greater
than the number of arcs in $G$ by $m_j-(m_i-1)= m_j-m_i+1>0$,
but this contradicts
the maximality of the number of arcs in
$G$. Hence each bicomponent in $G$, excluding perhaps one, contains one
vertex. Then the number of arcs in $G$ equals $(n-1)+(n-2)+...+(n-k+1)+
(n-k+1)(n-k)=n(n-k)+\sfrac{k^2-k}{2}$.
\rule{2mm}{2mm}

\bigskip

Since $\max\limits_{2\leq k\leq n}\left\{n(n-k)+\sfrac{k^2-k}{2} \right\}
=(n-1)^2$,
it follows from
Corollary \ref{cor12} that the number of arcs
in an $n$-vertex nonbiconnected digraph does not exceed $(n-1)^2$.

Let $D(n,k,d=\infty)$ denote the totality of all nonisomorphic
$n$-vertex $d$-critical digraphs of infinite diameter with $k$
bicomponents, and $\beta (n,k)=| D(n,k,d=\infty)|$.

\begin{Cor} \label{cor13}
\[
\beta(n,k)=\binom{n-1}{k-1} \hspace{7cm}
\]
\end{Cor}

{\bf Proof.\/}
Let us consider $k$ boxes numbered $1,...,k$ and $n$ identical points
contained in the boxes. There are no empty boxes. We consider the points
as vertices of a graph such that points in a box constitute a complete
symmetric digraph, and arcs go from all vertices in the boxes with
smaller numbers to all vertices in the boxes with greater numbers. There
are no other arcs in this graph. It follows from Theorem \ref{th1} that
distributing
in different ways $n$ identical points in $k$ boxes without empty ones and
constructing digraphs in the way as before, we can obtain all $d$-critical
digraphs of the totality $D(n,k,d=\infty)$.
All these digraphs are nonisomorphic. It was shown in \cite{A79} that the
number of ways of
distributing
$n$ identical objects in $k$ distinct boxes
without empty ones equals $\binom{n-1}{k-1}$.
\rule{2mm}{2mm}

\begin{Cor} \label{cor14}
The number of distinct $d$-critical digraphs of infinite diameter
with $k$ bicomponents that can be constructed on given $n$ numbered
vertices equals $k!\,S(n,k)$ where $S(n,k)$ are the Stirling numbers of
the second kind.
\end{Cor}

{\bf Proof.\/}
The proof of Corollary \ref{cor12} shows that the number of distinct
$d$-critical digraphs of infinite diameter with $k$ bicomponents
that can be constructed on $n$ numbered vertices equals the number of
ways of
distributing
$n$ distinct objects in $k$ distinct boxes without
empty ones. It was shown in \cite{A79} that the number equals $k!\,S(n,k)$.
\rule{2mm}{2mm}

\bigskip

Now we proceed to characterizing $d_m$-critical digraphs with
infinite quasi-diameter. As we have stated, in order to do that, it
suffices to characterize their Hertz graphs. Let $\Gamma$ be the Hertz graph
of a $d_m$-critical digraph with $d_m=\infty$. We noted in the
beginning of the section that the quasi-diameter of a digraph is infinite
iff there is a pair of vertices, neither of which is reachable from the
other. The above notes imply that $\Gamma$ is acyclic, transitive, and
has a pair of vertices, neither of which is reachable from the other.
We shall show that there exists exactly one pair of vertices in $\Gamma$
with this property. Indeed, suppose there exist two such pairs $\{a,b\}$
and $\{c,d\}$. Consider the subgraph of the graph $\Gamma$ induced by
the vertices $a,b,c,d$. It is easy to see that the subgraph is
isomorphic to either of the graphs depicted in Figures $1,2$.

\unitlength=1mm
\begin{figure}[hbpt]
\begin{picture}(150,50)

\put(20,20){\line(1,0){40}}               %
\put(20,20){\circle*{1}}                  %
\put(15,15){\makebox(5,5){$a$}}             %
\put(60,20){\circle*{1}}                  
\put(60,15){\makebox(5,5){$d$}}             %
\put(60,40){\circle*{1}}                  %
\put(60,40){\makebox(5,5){$c$}}             %
\put(40,5){\makebox(0,0){Fig. 1}}         %

\put(90,20){\line(1,0){40}}                  %
\put(90,40){\line(1,0){40}}                  %
\put(90,20){\line(2,1){40}}                  %
\put(90,40){\line(2,-1){40}}                 
\put(90,20){\circle*{1}}                     %
\put(85,15){\makebox(5,5){$a$}}                %
\put(90,40){\circle*{1}}                     %
\put(85,40){\makebox(5,5){$b$}}                %
\put(130,20){\circle*{1}}                    %
\put(130,15){\makebox(5,5){$d$}}               %
\put(130,40){\circle*{1}}                    %
\put(130,40){\makebox(5,5){$c$}}               %
\put(110,5){\makebox(0,0){Fig. 2}}           %

\thicklines
\put(40,20){\vector(1,0){0}}                  
\put(110,20){\vector(1,0){0}}                 
\put(110,40){\vector(1,0){0}}                 
\put(120,25){\vector(2,-1){0}}                
\put(120,35){\vector(2, 1){0}}                
\end{picture}
\end{figure}

Figure 1 corresponds to the case, when the sets $\{a,b\}$ and $\{c,d\}$
has a vertex in common; Figure 2 corresponds to the case, when the pairs
are disjoint. In both cases we add the arc $(c,d)$. It is easy to see
that the obtained graphs do not contain directed cycles, and in each graph
there exists a pair of vertices neither of which is reachable from the
other. Thus, we have shown that in the Hertz graph of a $d_m$-critical
digraph of infinite quasi-diameter there exists exactly one
pair of vertices neither of which is reachable from the other.

We denote by $\Gamma_{k,i}=(X_k,U_{k,i})$ a digraph such that
$$
X_k=\{1,...,k\}\mbox{ and } (s,j)\in U_{k,i} \iff
(s<j)\&[(s,j)\not=(i,i+1)]
$$
that is the graph $\Gamma_{k,i}$ is obtained from the graph $\Gamma_k$
by
removal of the arc $(i,i+1),
\linebreak
i\in \{1,...,k-1\}$.

\begin{Th} \label{th2}
Let $\Gamma$ be the Hertz graph of a $d_m$-critical digraph of
infinite quasi-diameter with $k$ bicomponents, $a$ and $b$ be the
vertices of $\Gamma$, neither of which is reachable from the other; the
indegree of the vertex $a$ equals $i$, $0\leq i\leq k-2$.
Then $\Gamma$ is isomorphic to the graph $\Gamma_{k,i+1}$.
\end{Th}

{\bf Proof.\/}
The above considerations show that the graph $\Gamma$ is acyclic,
transitive, and has the property that any two its vertices, excluding
the pair $\{a,b\}$, are adjacent. Hence the subgraph $\Gamma'$, that is
obtained from $\Gamma$ by
removal of the vertex $b$ and arcs
incident to it, is a complete acyclic transitive $(k-1)$-vertex digraph.
By Lemma \ref{lem1}, $\Gamma'$ is isomorphic to the graph $\Gamma_{k-1}$.
Let us renumber the vertices of $\Gamma'$ in the following way: as a
number of a vertex $x$ we consider the number $j$ that corresponds to
the vertex $x$ under the isomorphism $\Gamma' \leftrightarrow
\Gamma_{k-1}$ (evidently, such an isomorphism is unique). It is easy to
see that the number of the vertex $a$ is $i+1$. We shall show that
$\Gamma$ contains all arcs of the kind $(b,x_j)$ where $j>i+1$, and all
arcs of the kind $(x_t,b)$ where $t<i+1$. Indeed, if $\Gamma$ contained
the arc $(x_l,b)$ where $l>i+1$, then there would exist a path
$\{(x_{i+1},x_l),(x_l,b)\}$, $x_{i+1}=a$ from $a$ to $b$ contrary to the
hypothesis of mutual
unreachability
of the vertices $a$ and $b$. If
$\Gamma$ contained an arc of the kind $(b,x_m)$ where $m<i+1$, then
there would exist a path $\{(b,x_{m}),( x_{m}, x_{i+1})\}$ from
$b$ to $a$ contrary to the hypothesis again. But the vertex $b$ is
adjacent to each vertex in $\Gamma$, except the vertex $a$,
hence $\Gamma$ contains
all arcs of the kind $(b,x_j)$, $j>i+1$ and $(x_s,b)$, $s<i+1$. Now let
us add the arc $(a,b)$ to the graph $\Gamma$. One sees that the obtained
graph $\Gamma''$ is isomorphic to the graph $\Gamma_k$; under the
isomorphism the number $i+1$ corresponds to the vertex $a$, and $i+2$
corresponds to $b$. Hence $\Gamma$ is isomorphic to the graph that is
obtained from the graph $\Gamma_k$ by the removal of the arc
$(i+1,i+2)$.
\rule{2mm}{2mm}

\bigskip

From the notes on $d_m$-critical digraphs with $d_m=\infty$ one
sees that Theorem \ref{th2} gives the description of these digraphs up
to isomorphism.

By Corollary \ref{cor12} of Theorem \ref{th1}, the greatest number of
arcs in an $n$-vertex digraph of infinite diameter with $k$ bicomponents
equals $n(n-k)+\sfrac{k^2-k}{2}$. In view of the connection between
$d_m$-critical digraphs of infinite quasi-diameter and $d$-critical
digraphs of infinite diameter established by Theorems \ref{th1}
and \ref{th2} we obtain

\begin{Cor} \label{cor21}
If the number of arcs in an $n$-vertex digraph with $k\geq 3$
bicomponents is greater than $n(n-k)+\sfrac{k^2-k}{2}-1$, then its
quasi-diameter is finite.
\end{Cor}

It is easy to see that the number of arcs in an $n$-vertex digraph of
infinite quasi-diameter with two bicomponents does not exceed
$(n-1)(n-2)$. This result together with Corollary \ref{cor21} gives

\begin{Cor} \label{cor22}
If the number of arcs in an $n$-vertex digraph is greater than
$n^2-3n+2$, then its quasi-diameter is finite.
\end{Cor}

Let $q(n,k)$ denote the number of nonisomorphic $d_m$-critical
digraphs having $k$ bicomponents and infinite quasi-diameter; and
$q^*(n,k)$ denote the number of distinct $d_m$-critical
digraphs, having $k$ bicomponents and infinite quasi-diameter, that can
be constructed on $n$ numbered vertices.

\begin{Cor} \label{cor23}
\vspace*{-5mm}
$$
q(n,k)=\left\{
\begin{array}{ll}
\intt{\sfrac{n}{2}} & \mbox{ for }k=2,\\
(k-1)\displaystyle\sum\limits_{t=2}^{n-k+2}
\binom{n-t-1}{k-3}\intt{\sfrac{t}{2}}
& \mbox{ for }k>2.
\end{array}
\right.
$$
\end{Cor}

{\bf Proof.\/}
In the same manner as in the proof of Corollary \ref{cor13} to Theorem
\ref{th1} we consider $k$ boxes numbered $1,...,k$ and $n$ identical
points, contained in these boxes without empty ones. We consider the
points as vertices of a digraph such that the points in a box constitute
a complete symmetric graph, and arcs go from all points in the box
numbered $i$ to all points in the box numbered $j$ where $i<j$ and
$j\ne 2$.

The number of nonisomorphic digraphs obtained in different
distributions
of $n$ identical points in given $k$ boxes for $k>2$ equals
$$
\displaystyle\sum\limits_{t=2}^{n-k+2}
\binom{n-t-1}{k-3}\intt{\sfrac{t}{2}}.
$$
Each summand in the sum corresponds to the number of ways of %
distributing
$t$ points in two identical boxes, and the rest $n-t$ points in $k-2$
distinct boxes. Thus, we have counted the number of nonisomorphic
$d_m$-critical digraphs, whose Hertz graphs are isomorphic to
$\Gamma_{k,1}$. Evidently, for any $i$ and $j$ the number of
nonisomorphic $d_m$-critical digraphs, whose Hertz graph is
isomorphic to $\Gamma_{k,i}$, equals the number of nonisomorphic
digraphs, whose Hertz graph is isomorphic to $\Gamma_{k,j}$, therefore
the number of nonisomorphic $n$-vertex $d_m$-critical digraphs,
having $k$ bicomponents and infinite quasi-diameter, for $k>2$ equals
$$
(k-1)\displaystyle\sum\limits_{t=2}^{n-k+2}
\binom{n-t-1}{k-3}\intt{\sfrac{t}{2}}.
$$
If $k=2$ then the number of such digraphs is $\intt{\frac{n}{2}}$.
\rule{2mm}{2mm}

\begin{Cor} \label{cor24}
$$
q^*(n,k)=\left\{
\begin{array}{ll}
2^{n-1}-1 & \mbox{ for }k=2,\\
(k-1)\displaystyle\sum\limits_{t=2}^{n-k+2}
\binom{n}{t} (k-2)! \,S(n-t,k-2)(2^{t-1}-1)
& \mbox{ for }k>2.
\end{array}
\right.
$$
\end{Cor}

The proof of the corollary is analogous to the proof of Corollary
\ref{cor14} to Theorem \ref{th1}.

\bigskip
\begin{center}
\large \bf
2.
On critical digraphs with infinite values \\
  of radius and quasi-radius
\end{center}

In this section we shall characterize
up to isomorphism critical digraphs
with infinite values of radius and quasi-radius.

It is easy to see that $r$-critical and $r_m$-critical
digraphs with infinite values of $r$ and $r_m$, like $d$-critical
and $d_m$-critical digraphs with infinite values of $d$ and
$d_m$, are transitive. Hence the problem of describing these digraphs is
reduced to the problem of describing their Hertz graphs.

\bigskip

{\bf I. }In this subsection we shall characterize
$r$-critical digraphs
of infinite radius.

\begin{Th} \label{th3}
Let $\Gamma=(X,U)$ be the Hertz graph of a $r$-critical digraph $G$
with $k$ bicomponents and $r(G)=\infty$. Then the graph $\Gamma$
is isomorphic to the graph $\Gamma_{k,1}$.
\end{Th}

{\bf Proof.\/}
Lemma \ref{lem1} implies that the graph $\Gamma$ contains a source which
is a vertex that has no incoming arcs. Let $z$ denote the source. We
shall show that arcs go from the vertex $z$ to all but one vertices in
$\Gamma$. Indeed, assume there is a pair of vertices $u$ and $v$ that
are not reachable from $z$. Since the graph $\Gamma$ is antisymmetric,
at least one of the arcs $(u,v)$, $(v,u)$ is missing in it. Let for
instance, the arc $(u,v)$ be missing. Having added the arc $(z,u)$ we
obtain a graph $\Gamma'$ in which there are no directed cycles and whose
radius is infinite. It is impossible, since $\Gamma$ was a $r$-critical
digraph. Thus, arcs go from the vertex $z$ to all vertices in $\Gamma$,
except a vertex $y$.
Obviously, the vertex $y$ is also a source in
$\Gamma$. Hence arcs go from it to all vertices in $\Gamma$, except $z$.
Denote by $X'$  the set $X\setminus \{z,y\}$. It is easy to see that the
addition of an arc, both endpoints of which belong to $X'$, does not
decrease the radius. Since $\Gamma$ is a $r$-critical digraph,
the addition of an arc, both endpoints of which belong to $X'$, must
result in the appearance of a directed cycle. Since the subgraph
of  $\Gamma$ induced by the set $X'$ is transitive and antisymmetric,
what has been said and Lemma \ref{lem1} imply that this subgraph is a
$(k-2)$-vertex transitive tournament. This means that $\Gamma$ is
isomorphic to the graph $\Gamma_{k,1}$.
\rule{2mm}{2mm}

\bigskip

Theorem \ref{th3} is actually contained in the work \cite{I70} in a
somewhat different form.

We shall cite a few corollaries of Theorem \ref{th3}.

\begin{Cor} [\rm See also \cite{I70}] \label{cor31}
If an $n$-vertex digraph $G$ with $k>1$ bicomponents contains more than
$\lambda(n,k)$ arcs, then $r(G)<\infty$ where
$$
\lambda(n,k)=\left\{
\begin{array}{ll}
(n-1)(n-2), & \mbox{ if }k=2,\\
n(n-k)+\sfrac{k^2-k-2}{2},
& \mbox{ if }k>2.
\end{array}
\right.
$$
\end{Cor}

{\bf Proof.\/}
Let $G$ be an $n$-vertex digraph of infinite radius with $k$
bicomponents. Let $G$ contain the greatest number of arcs. Evidently, $G$
is a $r$-critical graph. Hence its Hertz graph is isomorphic to the graph
$\Gamma_{k,1}$. If $k=2$ then the number of arcs in an
$n$-vertex digraph, whose Hertz graph is isomorphic to $\Gamma_{2,1}$,
is maximal when one of its bicomponents contains $n-1$ vertices. Hence
in this case the digraph $G$ contains $(n-1)(n-2)$ arcs. Now let
$k\ge 3$. We shall show that the number of arcs in a digraph, whose
Hertz graph is isomorphic to $\Gamma_{k,1}$, is maximal in the case when
all its bicomponents, excluding one bicomponent not corresponding to the
vertices 1 and 2 of $\Gamma_{k,1}$, are one-vertex. Indeed, let
$2<i<j\le k$ and the number of vertices in the bicomponents,
corresponding to the vertices $i$ and $j$ of the graph $\Gamma_{k,1}$,
equal $k_i\ge 2$ and $k_j\ge 2$, respectively. Let us move all vertices
but one from the component corresponding to the vertex $j$ to the
component corresponding to the vertex $i$. The number of arcs,
connecting vertices of these bicomponents to vertices of other
bicomponents, evidently does not change. The number of arcs in the
subgraph induced by the vertices of these bicomponents was
$k_i(k_i-1)+k_j(k_j-1)+k_ik_j$, and became
$(k_i+k_j-1)(k_i+k_j-2)+(k_i+k_j-1)$; that is it has decreased by
$k_ik_j-k_i-k_j+1\ge 1$. It is easy to see that if we move all vertices
but one from each bicomponent, corresponding to the vertices 1 or 2 of
$\Gamma_{k,1}$, to the bicomponent, corresponding to the vertex $k$,
then the number of arcs in the obtained $r$-critical digraph will not
decrease. Thus, we have shown that the number of arcs in a $r$-critical
digraph whose Hertz graph is isomorphic to $\Gamma_{k,1}$ is
maximal, when all its bicomponents, excluding one not corresponding to
the vertices 1 and 2 of the graph $\Gamma_{k,1}$, are one-vertex. The
number of arcs in such an $n$-vertex digraph equals
$$
2(n-2)+(n-3)+...+(n-k+1)+(n-k+1)(n-k)=n(n-k)+\sfrac{k^2-k-2}{2},
$$
as required.
\rule{2mm}{2mm}

\begin{Cor} \label{cor32}
If a digraph $G$ has the property that $r(G)=\infty$ and the addition of
an arbitrary missing arc converts it into a digraph with finite
radius, then the Hertz graph of $G$ is isomorphic either to
$\Gamma_{3,1}$ or to $\Gamma_{2,1}$.
\end{Cor}

{\bf Proof.\/}
If the digraph $G$ has the property stated in Corollary \ref{cor32},
then so does its Hertz graph $\Gamma$. By Theorem \ref{th3}, the graph
$\Gamma$ is isomorphic to the graph $\Gamma_{k,1}$ for some $k$. If
$k\ge 4$ then the addition of the arc $(4,3)$ to $\Gamma_{k,1}$ does not
convert it into a digraph with finite radius. Hence $\Gamma$ is
isomorphic either to $\Gamma_{3,1}$ or to $\Gamma_{2,1}$. It is easy to
see that these graphs
possess
the property stated in Corollary
\ref{cor32}.
\rule{2mm}{2mm}

\begin{Cor} \label{cor33}
If an $n$-vertex digraph $G$ contains more than $(n-1)(n-2)$ arcs, then
$r(G)<\infty$.
\end{Cor}

{\bf Proof.\/}
If the number of bicomponents in the graph $G$ equals two, then the
number of its arcs evidently does not exceed $(n-1)(n-2)$, what
corresponds to the case, when either of the bicomponents is a complete
symmetric $(n-1)$-vertex graph, and the other is one-vertex. If the
number of bicomponents of $G$ is greater than two, then by Corollary
\ref{cor31}, the number of its arcs does not exceed
$n(n-k)+\sfrac{k^2-k-2}{2}$. The last expression does not exceed
$(n-1)(n-2)$, if $k\ge 3$.
\rule{2mm}{2mm}

\begin{Cor} \label{cor34}
The number of nonisomorphic $r$-critical $n$-vertex digraphs with $k$
bicomponents and infinite radius equals
$$
\nu (n,k)=\left\{
\begin{array}{ll}
\intt{\sfrac{n}{2}} & \mbox{ for }k=2,\\
\displaystyle\sum\limits_{t=2}^{n-k+2}
\intt{\sfrac{t}{2}} \binom{n-t-1}{k-3}
& \mbox{ for }k\ge 3.
\end{array}
\right.
$$
\end{Cor}

{\bf Proof.\/}
The Hertz graph of a critical digraph with $k$ bicomponents and infinite
radius is isomorphic to $\Gamma_{k,1}$. Let the bicomponents,
corresponding to the vertices 1 and 2 of $\Gamma_{k,1}$, contain
together $t$ vertices of a critical digraph and $k\ge 3$. Fix the number
of vertices in each of the bicomponents, corresponding to the vertices 1
and 2 of $\Gamma_{k,1}$.
Distributing
the rest $n-t$ identical vertices in
$k-2$ distinct bicomponents we obtain (see \cite{A79}) $\binom{n-t-1}{k-3}$
nonisomorphic critical digraphs. There are $\intt{\sfrac{t}{2}}$ ways of
distributing
$t$ vertices in bicomponents, corresponding to the
vertices 1 and 2 of $\Gamma_{k,1}$. Hence the number of nonisomorphic
$r$-critical digraphs of infinite radius having $n$ vertices, $k$
bicomponents, and such that the total number of vertices in
bicomponents, corresponding to the vertices 1 and 2 of $\Gamma_{k,1}$,
equals $t$ is $\intt{\sfrac{t}{2}} \binom{n-t-1}{k-3}$. Summing the
quantity over $t$ we obtain what we needed to prove. The proof in the
case $k=2$ is obvious.
\rule{2mm}{2mm}

\begin{Cor} \label{cor35}
The number of distinct $r$-critical digraphs with $k$ bicomponents
and infinite radius that can be constructed on $n$ numbered vertices
equals
$$
\nu^*(n,k)=\left\{
\begin{array}{ll}
2^{n-1}-1 & \mbox{ for }k=2,\\
\displaystyle\sum\limits_{t=2}^{n-k+2}
\binom{n}{t} (2^{t-1}-1)(k-2)! \,S(n-t,k-2)
& \mbox{ for }k>2.
\end{array}
\right.
$$
\end{Cor}

The proof of Corollary \ref{cor35} is analogous to the proof of
Corollary \ref{cor34}, but in this case we should consider the vertices
to be distinct.

\bigskip

{\bf II.} In this subsection we shall take up the study of $r_m$-critical
digraphs with infinite quasi-radius. As above, we shall
characterize
the structure of corresponding Hertz graphs.

We call a vertex in a digraph $G=(X,U)$ a {\em quasi-center\/}, if the
relation $\max\limits_{y\in X} \rho_m(x,y)<\infty$ holds.
Evidently, the quasi-radius of a digraph is finite iff the digraph
contains a quasi-center.

\begin{Lem} \label{lem2}
Let a $k$-vertex graph $\Gamma$ be the Hertz graph of a $r_m$-critical
digraph with infinite quasi-radius. In addition, suppose the
graph $\Gamma$ is not weakly connected. Then $\Gamma$ is isomorphic to
the graph $\Gamma_{k,0}=(X_{k,0},U_{k,0})$, where
$$
X_{k,0}=\{1,...,k\} \mbox{ and } (i,j)\in U_{k,0}\iff (i<j)\&(i\ne 1).
$$
\end{Lem}

{\bf Proof.\/}
Evidently, the graph $\Gamma$ can not contain more than two weak
components, since the quasi-radius of a graph containing more
than one weak component can not be finite. Let $X_1$ and
$X_2$ be the weak components of the graph $\Gamma$. If we
add an arc such that both its endpoints belong to one weak
component, then the quasi-radius of the obtained digraph remains
infinite. Hence the number of bicomponents must decrease. Then using
Lemma \ref{lem1}, we obtain that the subgraphs induced by the sets
$X_1$ and $X_2$ are transitive tournaments. If $\min\{|X_1|,|X_2|\}=1$
then  $\Gamma$ is isomorphic to the graph $\Gamma_{k,0}$. Let $|X_1|=l$,
$|X_2|=t$, and $\min\{l,t\}\ge 2$. Denote by $\Gamma^1$ and $\Gamma^2$
the subgraphs of $\Gamma$ induced by the sets $X_1$ and $X_2$,
respectively. Renumber the vertices in $\Gamma$ so, that the mapping
$\varphi:x_i\leftrightarrow i,\ i=1,...,l$ be an isomorphism of the graphs
$\Gamma^1$ and $\Gamma_l$, and the mapping
$\psi:x_{j+l}\leftrightarrow j,\ j=1,...,t$ be an isomorphism of the
graphs $\Gamma^2$ and $\Gamma_t$. According to the above considerations
it is possible. Let us add the arc $(x_1,x_{l+t})$ to the graph
$\Gamma$. Denote the obtained graph by $\Gamma'$. It is easy to see that
the number of bicomponents does not change. We shall show that the
quasi-radius of the obtained graph is infinite. Indeed, the vertex $x_l$
and each vertex $x_j,\,j>l$ are mutually unreachable. Hence no vertex
in the subgraph $\Gamma^2$ can be a quasi-center in the obtained graph.
On the other hand, the vertex $x_{l+1}$ and each vertex $x_i,\,i\le l$
are mutually unreachable. Hence none of the vertices $x_i,\,i\le l$ can
be a quasi-center in the graph $\Gamma'$. Thus we have proved that
$\Gamma'$ has no quasi-centers, hence its quasi-radius is infinite. This
implies that $\Gamma$ is not $r_m$-critical contrary to the
hypothesis.
\rule{2mm}{2mm}

\begin{Lem} \label{lem3}
Let $\Gamma$ be the Hertz graph of a $d_m$-critical digraph
with $r_m=\infty$. Let $\Gamma$ be weakly connected and each vertex in
$\Gamma$ have either outdegree or indegree equal to
zero. Then $\Gamma$ is isomorphic to the graph $D_4=(Y_4,V_4)$ where
$$
Y_4=\{1,2,3,4\} \mbox{ and } V_4=\{(1,3),(1,4),(2,3),(2,4)\}.
$$
\end{Lem}

{\bf Proof.\/}
Let $X_1$ be the totality of the vertices of the graph $\Gamma$ whose
indegree equals zero; $X_2$ be the totality of the vertices
of $\Gamma$ whose outdegree equals zero. Evidently, the sets
$X_1$ and $X_2$ are disjoint. Since $r(\Gamma)\ge r_m(\Gamma)=\infty$,
it follows from Theorem \ref{th3} that $|X_1|\ge 2$; since otherwise,
there exists a vertex in $X_2$, whose total degree equal zero contrary
to the weak
connectivity
of $\Gamma$. Suppose $|X_2|=1$. Since arcs
from $X_1$ can go only to $X_2$; either $r_m(\Gamma)=1$, if an arc goes
from each vertex of $X_1$ to the only vertex of $X_2$, or there exists a
vertex in $X_1$, whose total degree equals zero, contrary to the
assumption of $\Gamma$ being weakly connected. Hence $|X_2|\ge 2$. We
shall show that $|X_1|=|X_2|=2$. Let, for instance, $|X_1|\ge 3$. Add an
arc, both endpoints of which belong to $X_1$. Denote the obtained graph
by $\Gamma'$. Evidently, the number of bicomponents in $\Gamma'$ equals
the number of bicomponents in the digraph $\Gamma$. We shall show that
the quasi-radius of $\Gamma'$ is infinite. Indeed, since $|X_1|\ge 3$,
for any vertex $x\in X_1$ there exists a vertex $y\in X_1$ such that
$\rho_m(x,y)=\infty$. Hence $X_1$ contains no quasi-centers of
$\Gamma'$. The set $X_2$ contains no quasi-center of $\Gamma'$ either,
because any two vertices in $X_2$ are mutually unreachable and
$|X_2|\ge 2$. Hence, if $|X_1|\ge 3$, then the graph $\Gamma$ is not
$r_m$-critical. Evidently, an analogous argument is valid for the
set $|X_2|$. Thus, we have shown that $|X_1|=|X_2|=2$. It means that
$\Gamma$ is isomorphic to a subgraph of the graph $D_4$. However, the
graph $D_4$ has no directed cycles, and $r_m(D_4)=\infty$. Hence
$\Gamma\cong D_4$. On the other hand, it is easy to see that the graph
$D_4$ is $r_m$-critical.
\rule{2mm}{2mm}

\begin{Lem}[\bf basic] \label{lem4}
Let $\Gamma$ be the Hertz graph of a weakly connected $r_m$-critical
digraph with $r_m=\infty$. Then the vertex set of the graph
$\Gamma$ can be partitioned into two disjoint subsets such that

a) the subgraphs induced by each of these subsets are $r_m$-critical
with $r_m=\infty$;

b) an arc goes from each vertex of the first subset to each vertex of
the second subset.
\end{Lem}

In order to prove Lemma \ref{lem4} we need a few lemmas.

If each vertex in a graph $\Gamma$ has either outdegree or
indegree equal to zero, then by Lemma \ref{lem3}, $\Gamma$ is
isomorphic to the graph $D_4$; and the graph $D_4$ satisfies Lemma
\ref{lem4}.

So, let $\Gamma$ contain a vertex $v$ such that outdegree and
indegree of the vertex are greater than zero.  Denote by
$B_v$ the set of vertices of $\Gamma$ reachable from $v$, by $A_v$ the
set of vertices of $\Gamma$ from which the vertex $v$ is reachable, and
by $C_v$ the totality of all other vertices. Denote by $\Gamma(A_v)$,
$\Gamma(B_v)$, and $\Gamma(C_v)$ the subgraphs of the graph $\Gamma$
that are induced by the sets $A_v$, $B_v$, and $C_v$, respectively.
Since the graph $\Gamma$ is transitive, an arc goes from each vertex of
the set $A_v$ to each vertex of the set $B_v\cup\{v\}$, and from the
vertex $v$ to each vertex of the set $B_v$.
Obviously, $C_v\ne\emptyset$
since otherwise, the vertex $v$ would be a quasi-center in $\Gamma$.

\begin{Lem} \label{lem5}
Let $\Gamma$ be the Hertz graph of a $r_m$-critical digraph with
infinite quasi-radius; $A_v, B_v, C_v,\{v\}$ be the sets defined above.
Suppose $r_m(\Gamma(A_v))= r_m(\Gamma(B_v))=\infty$. Then

1) An arc goes from each vertex of the set $A_v$ to each vertex of the
set $C_v$; an arc goes from each vertex of the set $C_v$ to each vertex
of the set $B_v$.

2) The subgraph $\Gamma(C_v)$ is a transitive tournament.

3) The subgraphs $\Gamma(A_v)$ and $\Gamma(B_v)$ are $r_m$-critical.
\end{Lem}

{\bf Proof.\/}
First we shall show that $\Gamma(A_v)$ and $\Gamma(B_v)$ are
$r_m$-critical. Let us add an arc, both endpoints of which belong
to $B_v$, to the graph $\Gamma$. Denote the obtained graph by
$\Gamma'$. No vertex of the set $C_v$ can be a quasi-center in
$\Gamma'$, for each of such vertices and the vertex $v$ are mutually
unreachable, the vertex $v$ can not be a quasi-center of $\Gamma'$ for
the same reason. To any vertex in the set $A_v$ there exists a vertex in
this set such that these two vertices are mutually unreachable. Hence $A_v$
contains no quasi-centers of $\Gamma'$. Since $\Gamma'$ is $r_m$-critical
and $r_m(\Gamma)=\infty$, it follows that either $B_v$ contains
a quasi-center of $\Gamma'$ that is also a quasi-center in the graph
$\Gamma(B_v)$, or $\Gamma'$ contains a directed cycle. However, the addition
of an arc, both endpoints of which belong to $B_v$, can result in the
appearance of directed cycle only in the subgraph $\Gamma(B_v)$. This
consideration shows that $\Gamma(B_v)$ is $r_m$-critical. An
analogous consideration is evidently valid with respect to the subgraph
$\Gamma(A_v)$. Thus, statement 3) of the lemma is proved.

Now let us prove statement 2). Add to $\Gamma$ all missing arcs that go
from $A_v$ to $C_v$ and from $C_v$ to $B_v$. Denote the obtained graph
by $\Gamma''$. Evidently the number of bicomponents of this graph equals
the number of bicomponents of the graph $\Gamma$. The same
considerations as in the proof of statement 3) show that quasi-radius of
the graph $\Gamma''$ is infinite. Finally, let us show the graph
$\Gamma(C_v)$ is a transitive tournament. Indeed, the addition of an
arc, both endpoints of which belong to $C_v$, can not result in the
appearance of a quasi-center in the obtained graph, since each vertex in
$C_v$ and the vertex $v$ are mutually unreachable, and
$r_m(\Gamma(A_v))= r_m(\Gamma(B_v))=\infty$. Hence it must result in the
appearance of a directed cycle. However the subgraph $\Gamma(C_v)$ is the
only subgraph in which a directed cycle can appear. It means that the
transitive
antisymmetric digraph $\Gamma(C_v)$ has the property that the addition
of an arbitrary arc missing in it results in the appearance of a
directed cycle. Thus, it follow from Lemma \ref{lem1} that this digraph is a
transitive tournament.
\rule{2mm}{2mm}

\begin{Lem} \label{lem6}
Let $\Gamma$ be the Hertz graph of a $r_m$-critical digraph with
$r_m=\infty$. If $\min\{r_m(\Gamma(A_v)), r_m(\Gamma(B_v))\}<\infty$,
then $|C_v|=1$.
\end{Lem}

{\bf Proof.\/}
Let, for instance, $r_m(\Gamma(A_v))<\infty$.

We shall show first that the set $C_v$ contains a vertex that is
reachable from no quasi-center of the graph $\Gamma(A_v)$. Suppose
conversely that any vertex of the set $C_v$ is reachable from some
quasi-center of the graph $\Gamma(A_v)$. Denote by $Z_v$ the totality of
all quasi-centers of the graph $\Gamma(A_v)$. It is easy to see that the
subgraph $\Gamma(Z_v)$ of the graph $\Gamma(A_v)$, induced by the set
$Z_v$, is a transitive tournament. Let a vertex $z_0\in Z_v$ correspond
to the vertex 1 under the isomorphism $\Gamma(Z_v)\leftrightarrow
\Gamma_{|Z_v|}$. Then all vertices of the set $Z_v$ are reachable from
$z_0$ and, by assumption, so are all vertices of the set $C_v$. But in
this case the vertex $z_0$ is a quasi-center, what is impossible. Thus
we have shown that $C_v$ contains a vertex that is reachable from no
quasi-center of the graph $\Gamma(A_v)$. Now we shall show that the set
$C_v$ contains exactly one vertex with the above property.

Suppose there exist two such vertices $y_1$ and $y_2$; suppose the arc
$(y_2,y_1)$ is missing in the graph $\Gamma(C_v)$. Let us add an arc to
$\Gamma$ that goes from an arbitrary quasi-center of $\Gamma(A_v)$  to
the vertex $y_2$. The following consideration shows that we shall obtain
a graph $\Gamma^1$ with infinite quasi-radius.

No vertex in the set $C_v$ can be a quasi-center in $\Gamma^1$, since
such a vertex and the vertex $v$ are mutually unreachable, hence the
vertex $v$ can not be a quasi-center in $\Gamma^1$ either. None of the
quasi-centers of the graph $\Gamma(A_v)$ can be a quasi-center in
$\Gamma^1$, since each of them and the vertex $y_1$ are mutually
unreachable; no other vertex of the set $A_v$ can be a quasi-center in
the graph $\Gamma^1$, since no such vertex is a quasi-center in the
graph $\Gamma(A_v)$. Evidently, the set $B_v$ contains no quasi-centers.

Let $y\in C_v$ be that unique vertex that is reachable from no
quasi-center of the graph $\Gamma(A_v)$. Suppose, in addition,
$C_v\backslash \{y\}\ne \emptyset$. It is easy to see that the vertex
$y$ is reachable from no vertex of the set $C_v\setminus \{y\}$.

Consider next two cases.

{\it (i)\/} $r_m(\Gamma(B_v))=\infty$. Let us add the arc $(y,v)$ to the
graph $\Gamma$. We shall show that quasi-radius of the obtained graph
$\Gamma^2$ is infinite. The vertex $y$ remains mutually reachable with
no quasi-center of the graph $\Gamma(A_v)$, hence neither vertex $y$ nor
any vertex in $A_v$ can be a quasi-center in the graph $\Gamma^2$.
No vertex of the set $C_v\backslash \{y\}$ is mutually reachable with
the vertex $v$, hence no such vertex can be a quasi-center in the graph
$\Gamma^2$. No vertex of the set $B_v$ can be a quasi-center in the
graph $\Gamma^2$ for the reason that $r_m(\Gamma(B_v))=\infty$.

{\it (ii)\/} $r_m(\Gamma(B_v))<\infty$. The above considerations show
that if $r_m(\Gamma(A_v))<\infty$ then the set $C_v$ contains exactly
one vertex $y$ that is mutually reachable with no quasi-center of the
graph $\Gamma(A_v)$. An analogous consideration shows that if
$r_m(\Gamma(B_v))<\infty$ then the graph $\Gamma(C_v)$ contains
exactly one vertex $u$ that is mutually reachable with no quasi-center of
the graph $\Gamma(B_v)$. Here evidently no vertex in the set
$C_v\setminus \{u\}$ is reachable from the vertex $u$. Suppose first
that $y=u$. Then by the above consideration, it is an isolated vertex in
the graph $\Gamma(C_v)$. Let us add the arc $(t,v)$ where
$t\in C_v\setminus \{y\}$. We shall show that quasi-radius of the
obtained graph $\Gamma^3$ is infinite.

No vertex in the set $C_v\cup \{v\}$ is a quasi-center in $\Gamma^3$,
since the vertex $y$ is mutually reachable with no vertex in the set
$(C_v\cup \{v\})\setminus\{y\}$. No vertex in the set $A_v\cup B_v$ can
be a quasi-center in the graph $\Gamma^3$ for the reason that the vertex
$y$ is mutually reachable with no quasi-center in the graph
$\Gamma(A_v)$ and $\Gamma(B_v)$.

Suppose now $y\ne u$. Then by the above considerations, the arc $(u,y)$
is missing in the graph $\Gamma(C_v)$. Let us add the arc $(v,u)$ to the
graph $\Gamma$ and verify that quasi-center of the obtained graph
$\Gamma^4$ is infinite. Taking into account that the arc $(u,y)$ is missing
in the graph $\Gamma(C_v)$ we see that no vertex in the set $A_v$ can be a
quasi-center in the graph $\Gamma^4$, since the vertex $y$ is mutually
reachable with no quasi-center in the graph $\Gamma(A_v)$. No vertex in
the set $B_v\cup \{u\}$ can be a quasi-center in the graph $\Gamma^4$,
since the vertex $u$ is mutually reachable with no quasi-center of the
graph $\Gamma(B_v)$. And no vertex of the set $(C_v\cup
\{v\})\setminus\{u\}$ can be a quasi-center in $\Gamma^4$ for the reason
that the vertex $v$ is mutually reachable with no vertex in the set
$C_v \setminus \{u\}$. Thus, in the cases
$\{r_m(\Gamma(A_v))<\infty\}\&\{r_m(\Gamma(B_v))=\infty\}$ and
$\{r_m(\Gamma(A_v))<\infty\}\&\{r_m(\Gamma(B_v))<\infty\}$ the
assumption $|C_v|\ge 2$ leads to a contradiction.

Let us show that the only unconsidered case
$\{r_m(\Gamma(A_v))=\infty\}\&\{r_m(\Gamma(B_v))<\infty\}$ can be
reduced to case {\it (i)\/}. We note that in view of the symmetry of the
function $\rho_m$ the graph $\Gamma$ is $r_m$-critical iff so is
the graph $\mybar{\Gamma}$ that is obtained by reversing
the direction of the arcs in the graph $\Gamma$. Let $\mybar{A}_v$ be
the totality of vertices of the graph $\mybar{\Gamma}$ from which the
vertex $v$ is reachable; $\mybar{B}_v$ be the totality of vertices of the
graph $\mybar{\Gamma}$ reachable from the vertex $v$; $\mybar{C}_v$ be
the totality of all other vertices of the graph $\mybar{\Gamma}$.
It is obvious that
$$
\mybar{A}_v=B_v,\ \mybar{B}_v=A_v,\ \mybar{C}_v=C_v;\quad
r_m(\mybar{\Gamma}(\mybar{A}_v))=r_m(\Gamma(B_v)),\
r_m(\mybar{\Gamma}(\mybar{B}_v))=r_m(\Gamma(A_v)).
$$
Suppose that $r_m(\Gamma(A_v)) \! = \! \infty$ and
$r_m(\Gamma(B_v)) \! < \! \infty$.
Hence $r_m(\mybar{\Gamma}(\mybar{B}_v)) \! = \! \infty$ and
$r_m(\mybar{\Gamma}(\mybar{A}_v)) \! < \! \infty$.

Then case {\it (i)\/} considered above implies that
$|\mybar{C}_v|=|C_v|=1$.
\rule{2mm}{2mm}

\begin{Lem} \label{lem7}
Let $\min\{r_m(\Gamma(A_v)), r_m(\Gamma(B_v))\}<\infty$, $C_v=\{y\}$.
Denote by $A_y$ the set of vertices of the graph $\Gamma$ from which the
vertex $y$ is reachable; by $B_y$ the set of vertices reachable from
$y$; by $C_y$ the totality of all other vertices. Denote by
$\Gamma(A_y), \Gamma(B_y), \Gamma(C_y)$ the subgraphs of the graph
$\Gamma$ induced by the sets $A_y, B_y, C_y$, respectively. Then the
graph $\Gamma$ satisfies one of the conditions listed below:

1) $A_y\ne \emptyset,\ B_y\ne \emptyset,\ r_m(A_y)=r_m(B_y)=\infty$ and
the graph $\Gamma$ is of the form indicated in Lemma \ref{lem5}.

2) $A_y=\emptyset,\ B_y\ne \emptyset,\ r_m(B_y)=\infty$, the subgraph
$\Gamma(B_y)$ is $r_m$-critical, the subgraph $\Gamma(C_y)$ is a
transitive tournament, and an arc goes from each vertex of the set $C_y$
to each vertex of the set $B_y$.

3) $A_y\ne\emptyset,\ B_y=\emptyset,\ r_m(A_y)=\infty$, the subgraph
$\Gamma(A_y)$ is $r_m$-critical, the subgraph $\Gamma(C_y)$ is a
transitive tournament, and an arc goes from each vertex of the set $A_y$
to each vertex of the set $C_y$.
\end{Lem}

{\bf Proof.\/}
First we note that $A_y\subseteq A_v$ and $B_y\subseteq B_v$. Now we set
about proving case 2 of the lemma.

Let $A_y=\emptyset,\ B_y\ne\emptyset$. Then the vertex $y$ is reachable
from no vertex of the set $A_v$. In this case the addition of an arc,
both endpoints of which belong to the set $A_v$, can not result in the
appearance of a quasi-center in the graph $\Gamma$; hence it must
decrease the number of bicomponents of the graph $\Gamma(A_v)$. Thus, we
conclude that the graph $\Gamma(A_v)$ is a transitive tournament.

Consider next two cases.

{\it (i)\/} Let $r_m(B_v)=\infty$. Then the addition of an arc that goes
from $y$ to some vertex of the set $B_v$ can result neither in the
appearance of a quasi-center in the graph $\Gamma$ nor in reducing the
number of its bicomponents. Hence $B_y=B_v$. But in this case
$C_y=A_v\cup\{v\}$. The facts that $\Gamma(A_v)$ is a transitive
tournament and an arc goes from each vertex of the set $A_v$ to the
vertex $v$ imply that $\Gamma(C_y)=\Gamma(A_v\cup\{v\})$ is a transitive
tournament, and the graph $\Gamma$ satisfies condition 2 of the lemma.

{\it (ii)\/} Let $r_m(B_v)<\infty$. Denote by $W_v$ the totality of
quasi-centers of the graph $\Gamma(B_v)$. The $W_v$ is a transitive
tournament. Let  $w$ be a vertex in $W_v$ that is reachable from each
vertex of the set $W_v$. Denote by $\mybar{W}_v$ the totality of
vertices of the set $B_v$ from which the vertex $w$ is reachable.
Evidently, no vertex in the set $\mybar{W}_v$ is adjacent to the vertex
$y$, since otherwise, the quasi-radius of the graph $\Gamma$ would be
finite. It follows from the transitivity of the graph $\Gamma(B_v)$ and
from the definition of the set $\mybar{W}_v$ that an arc goes from each
vertex of the set $\mybar{W}_v$ to each vertex of the set
$B_v\setminus\mybar{W}_v$ (hence there are no arcs from the set
$B_v\setminus\mybar{W}_v$ to the set $\mybar{W}_v$). It is easy to see
that the addition of an arbitrary arc, both endpoints of which belong to
$\mybar{W}_v$, can not result in the appearance of a quasi-center in the
graph $\Gamma$. Hence it must decrease the number of bicomponents  of
the graph $\Gamma(\mybar{W}_v)$. Hence $\Gamma(\mybar{W}_v)$ is a
transitive tournament; taking into account that an arc goes from each
vertex of $\mybar{W}_v$ to each vertex of $B_v\setminus\mybar{W}_v$,  we
obtain $\mybar{W}_v=W_v$. It is easy to see that an arc goes from the
vertex $y$ to each vertex of the set $B_v\setminus W_v$. Now let us
investigate the graph $\Gamma(B_v\setminus W_v)$. The quasi-radius of
the digraph is infinite; since otherwise, taking into account what has
been said, we would obtain that quasi-radius of $\Gamma$ is finite. On
the other hand, the addition of an arbitrary arc,
whose both endpoints
belong to $B_v\setminus W_v$, must result in the appearance of a
quasi-center in the graph $\Gamma$ or in
decreasing
the number of
bicomponents of the graph $\Gamma$. But there are no arcs that go from
the vertices of the set $B_v\setminus W_v$. Hence the addition of an
arbitrary arc, both endpoints of which belong to $B_v\setminus W_v$,
must result either in
decreasing
the number of bicomponents of the graph
$\Gamma(B_v\setminus W_v)$ or in the appearance of a quasi-center in it.
Hence the graph $\Gamma(B_v\setminus W_v)$ is $r_m$-critical with
infinite quasi-radius. In order to prove case 2 of the lemma, we just
have to note that $C_y=W_v\cup\{v\}\cup A_v$ and the graph
$\Gamma(C_y)=\Gamma(A_v\cup\{v\}\cup W_v)$ is a transitive tournament.

Case 3 of the lemma like in the proof of Lemma \ref{lem6} can be reduced
to case 2 with the help of considering the graph $\mybar{\Gamma}$.

It remains to consider case 1. Let $A_y\ne \emptyset,\ B_y\ne\emptyset$.
Let $Z_v$ be the totality of quasi-centers of the graph $\Gamma(A_v)$,
and $W_v$ be the totality of quasi-centers of the graph $\Gamma(B_v)$
(either of these sets can be empty, in this case the corresponding graph
$\Gamma(A_v)$ or $\Gamma(B_v)$ is $r_m$-critical with
$r_m=\infty$). Denote by $\mybar{Z}_v$ the totality of vertices of the
set $A_v$ that are reachable from at least one vertex of the set $Z_v$;
denote by $\mybar{W}_v$ the totality of vertices of the set $B_v$ that
are reachable from at least one vertex of the set $W_v$. It is easy to
show in the same manner as above that $\mybar{Z}_v=Z_v$ and
$\mybar{W}_v=W_v$. Thus, we obtain that an arc goes from each vertex of
the set $A_v\setminus Z_v$ to each vertex of the set $Z_v$, and an arc
goes from each vertex of the set $W_v$ to each vertex of the set
$B_v\setminus W_v$. Now taking into account that the sets $A_v\setminus
Z_v$ and $B_v\setminus W_v$ contain no quasi-centers of the graphs
$\Gamma(A_v)$ and $\Gamma(B_v)$ respectively; one obtains easily that
$A_y= A_v\setminus Z_v$ and $B_y=B_v\setminus W_v$. Here
$\Gamma(A_v\setminus Z_v)$ and $\Gamma(B_v\setminus W_v)$ are
$r_m$-critical digraphs with $r_m=\infty$, and $C_y=Z_v\cup\{v\}\cup W_v$.
But $\Gamma(Z_v)$ and $\Gamma(W_v)$ are transitive
tournaments, an arc goes from each vertex of the set $Z_v$ to the vertex
$v$, and from the vertex $v$ to each vertex of the set $W_v$. Hence
$\Gamma(Z_v\cup\{v\}\cup W_v)$ is a transitive tournament. Thus, we
obtain that $\Gamma(A_y)$ and $\Gamma(B_y)$ are $r_m$-critical
with $r_m=\infty$, $\Gamma(C_y)$ is a transitive tournament; and an arc
goes from each vertex of the set $A_y$ to each vertex of the set $C_y$,
and from each vertex of the set $C_y$ to each vertex of the set $B_y$.
\rule{2mm}{2mm}

\bigskip

{\bf Proof of Lemma \ref{lem4}.\/}
It follows from Lemmas $3,5,7$ that the Hertz graph of a weakly
connected $r_m$-critical digraph with infinite quasi-radius is
either isomorphic to the graph $D_4$ or is of the form represented in
Figures $3,4,5$.

\begin{figure}[hbpt]
\begin{picture}(150,70)

\put(15,45){\circle{13}}                  
\put(15,45){\makebox(0,0){$A_y$}}         %
\put(41,45){\circle{13}}                  %
\put(41,45){\makebox(0,0){$B_y$}}         %
\put(28,25){\circle{13}}                  %
\put(28,25){\makebox(0,0){$C_y$}}         %
\put(21.5,45){\line(1,0){13}}             
\put(15,51.5){\line(1,1){13}}             
\put(41,51.5){\line(-1,1){13}}            
\put(28,64.5){\circle*{1}}                %
\put(28,65){\makebox(5,3){$y$}}             %
\put(15,38.5){\line(1,-1){8.6}}           
\put(41,38.5){\line(-1,-1){8.6}}          
\put(28,10){\makebox(0,0){Fig. 3}}        %

\put(80,45){\circle{13}}                  
\put(80,45){\makebox(0,0){$B_y$}}         %
\put(67,25){\circle{13}}                  %
\put(67,25){\makebox(0,0){$C_y$}}         %
\put(67,64.5){\circle*{1}}                %
\put(67,65){\makebox(5,3){$y$}}             %
\put(80,51.5){\line(-1,1){13}}            
\put(80,38.5){\line(-1,-1){8.6}}          
\put(75,10){\makebox(0,0){Fig. 4}}        %

\put(110,45){\circle{13}}                  
\put(110,45){\makebox(0,0){$A_y$}}         %
\put(123,25){\circle{13}}                  %
\put(123,25){\makebox(0,0){$C_y$}}         %
\put(123,64.5){\circle*{1}}                %
\put(123,65){\makebox(5,3){$y$}}             %
\put(110,51.5){\line(1,1){13}}             
\put(110,38.5){\line(1,-1){8.6}}           
\put(118,10){\makebox(0,0){Fig. 5}}        %

\put(15,38.5){\line(1,-1){8.6}}           
\put(41,38.5){\line(-1,-1){8.6}}          
\thicklines

\put(30,45){\vector(1,0){0}}                  
\put(21.5,58){\vector(1,1){0}}                
\put(35.5,57){\vector(1,-1){0}}               
\put(20.3,33.2){\vector(1,-1){0}}             
\put(37.7,35.2){\vector(1,1){0}}              

\put(116.5,58){\vector(1,1){0}}                  
\put(115.5,33.2){\vector(1,-1){0}}               
\put(74.5,57){\vector(1,-1){0}}                  
\put(76.5,35.2){\vector(1,1){0}}                 
\end{picture}
\end{figure}

Here the graphs $\Gamma(A_y)$ and $\Gamma(B_y)$ are $r_m$-critical
with $r_m=\infty$, the graph $\Gamma(C_y)$ is a transitive
tournament; the arrows mean that arcs go from each vertex of the set, from
which an arrow issues, to each vertex of the set, which the arrow enter.
The decomposition of the vertex set of the graph $D_4$ into the subsets
$\{1,2\}$ and $\{3,4\}$ evidently satisfies the assertion of Lemma
\ref{lem4}. Suppose the graph $\Gamma$ is of the form represented in
Figure 3. Consider the decomposition of the vertex set of the graph into
the subsets $A_y$ and $C_y\cup \{y\}\cup B_y$. The graph $\Gamma(A_y)$
is $r_m$-critical with $r_m=\infty$ and an arc goes from each
vertex of the set $A_y$ to each vertex of the set
$C_y\cup \{y\}\cup B_y$. We shall show that the graph
$\Gamma(C_y\cup \{y\}\cup B_y)$ is $r_m$-critical with
$r_m=\infty$.

It is obvious that $r_m(\Gamma(C_y\cup \{y\}\cup B_y))=\infty$. The
addition of an arbitrary arc, both endpoints of which belong to $B_y$,
results either in 
decreasing
the number of bicomponents or in the
appearance of a quasi-center. The addition of an arc going from $B_y$ to
$C_y\cup\{y\}$ decreases the number of bicomponents. The addition of an
arc connecting the vertex $y$ to some vertex of the set $C_y$ results in
a vertex of $C_y$ adjacent to $y$ becoming a quasi-center in the graph
$\Gamma(C_y\cup \{y\}\cup B_y)$. Hence the graph
$\Gamma(C_y\cup \{y\}\cup B_y)$ is $r_m$-critical with
$r_m=\infty$.

Let us now consider the case corresponding to Figure 4. In this case the
desired decomposition is given by the sets $C_y\cup \{y\}$ and $B_y$;
since the graph $\Gamma(C_y\cup \{y\})$ is isomorphic to the graph
$\Gamma_{k,0}$ for some $k$, and hence it is $r_m$-critical with
$r_m=\infty$, and $B_y$ has the property by the hypothesis. The case
corresponding to Figure 5 can be reduced to the case corresponding to
Figure 4 by reversing the direction of the arcs.
\rule{2mm}{2mm}

\bigskip

Now we can proceed with proving the principal result of the subsection.

\begin{Th} \label{th4}
Let $\Gamma$ be the Hertz graph of a $r_m$-critical digraph with
infinite quasi-radius. Then the vertex set of the graph $\Gamma$ can be
partitioned into $s$ disjoint subsets $X_1,...,X_s$ for some $s$ such
that

{\em a)\/} $1\le s\le\intt{\sfrac{k}{2}}$ where $k$ is the number of
vertices in the graph $\Gamma$;

{\em b)\/} $|X_i|=k_i\ge 2$ and the graph $\Gamma(X_i)$ is isomorphic to
the graph $\Gamma_{k_i,0}$ for $i=1,...,s$;

{\em c)\/} an arc goes from each vertex of the set $X_i$ to each vertex
of the set $X_j$ for $1\le i<j\le s$;

{\em d)\/} there are no arc, besides those named, in the graph $\Gamma$.
\end{Th}

{\bf Proof.\/}
If the graph $\Gamma$ is not weakly connected, then by Lemma \ref{lem2},
it is isomorphic to the graph $\Gamma_{k,0}$. Hence it satisfies the
assertion of the theorem for $s=1$. Therefore in the sequel we may
confine ourselves to considering only weakly connected graphs.

First we shall show that the vertex set of the graph $\Gamma$ can be
partitioned into two subsets $X_1$ and $M_1$ such that the graph
$\Gamma(X_1)$ is isomorphic to the graph $\Gamma_{t,0}$ for some $t$,
the graph $\Gamma(M_1)$ is $r_m$-critical with $r_m=\infty$, and
an arc goes from each vertex of the set $X_1$ to each vertex of the set
$M_1$. By Lemma \ref{lem4}, the vertex set of the graph $\Gamma$ can be
partitioned into two subsets $Y_1$ and $N_1$ such that the graphs
$\Gamma(Y_1)$ and $\Gamma(N_1)$ are $r_m$-critical with
$r_m=\infty$ and an arc goes from each vertex of the set $Y_1$ to each
vertex of the set $N_1$. If the graph $\Gamma(Y_1)$ is not weakly
connected, then the assertion is proved; otherwise, applying Lemma
\ref{lem4} to the graph $\Gamma(Y_1)$ we obtain a decomposition of the
set $Y_1$ into subsets $Y_2$ and $N_2$ with corresponding properties. It
is easy to see that the graph $\Gamma(N_2\cup N_1)$ is $r_m$-critical
with $r_m=\infty$. We continue this procedure until at some
step $m$ we obtain a graph $\Gamma(Y_m)$ that is not weakly connected
and hence is isomorphic to the graph $\Gamma_{|Y_m|,0}$. It is easy to
show, by induction, that the graph $\Gamma(N_1\cup...\cup N_m)$ is
$r_m$-critical with $r_m=\infty$. Setting $X_1=Y_m$ and
$M_1=N_1\cup...\cup N_m$ completes the proof of the assertion.

Now we continue proving the theorem. If the graph $\Gamma(M_1)$ is not
weakly connected, then the graph $\Gamma(M_1)$ is isomorphic to the
graph $\Gamma_{|M_1|,0}$, and the theorem is proved. Otherwise, applying
the preceding assertion to the graph $\Gamma(M_1)$ we obtain a
decomposition of the vertex set of the graph $\Gamma$ into three subsets
$X_1,X_2,M_2$; here the graph $\Gamma(X_2)$ is isomorphic to the graph
$\Gamma_{|X_2|,0}$, the graph $\Gamma(M_2)$ is $r_m$-critical
with $r_m=\infty$, and an arc goes from each vertex of the set $X_1$ to
each vertex of the set $X_2\cup M_2$ and from each vertex of the set
$X_2$ to each vertex of the set $M_2$. Continue this process by
induction until at some step $s$ we obtain that the graph $\Gamma(M_s)$
is not weakly connected. Then setting $X_s=M_s$ we obtain that the
totality of sets $\{X_1,...,X_s\}$ is the desired decomposition of the
vertex set of the graph $\Gamma$.
\rule{2mm}{2mm}

\bigskip

Note that if we introduce a relation $T$  on the vertex set of the graph
$\Gamma$ in the following way:
$$
xTy\Leftrightarrow \mbox{both vertices $x$ and $y$ belong to one set in the
decomposition } \{X_1,...,X_s\};
$$
then it follows from Theorem \ref{th4} that the factor graph of the
graph $\Gamma$ relative to the relation $T$ is isomorphic to the graph
$\Gamma_s$.

Now we cite a few corollaries of Theorem \ref{th4}.

\begin{Cor} \label{cor41}
Let $G$ be an $n$-vertex digraph with $k\ge 2$ bicomponents. If the
number of arcs in $G$ is greater than $n(n-k-1)+\intt{\sfrac{k^2}{2}}$,
then $r_m(G)<\infty$.
\end{Cor}

{\bf Proof.\/}
Denote by $\Gamma_{k,s;k_1,...,k_s}$ a $k$-vertex digraph, whose vertex
set is partitioned into $s$ nonintersecting classes $Y_1,...,Y_s$,
$|Y_\alpha|=k_\alpha\ge 2$, $\alpha=1,...,s$; the subgraph induced by
the set $Y_\alpha$ is isomorphic to $\Gamma_{k_\alpha,0}$, and, in
addition, if $\alpha_1<\alpha_2$ then an arc goes from each vertex of
the class $Y_{\alpha_1}$ to each vertex of the class $Y_{\alpha_2}$;
there are no other arcs in the digraph $\Gamma_{k,s;k_1,...,k_s}$.

Obviously, an $n$-vertex digraph $D$ of infinite quasi-radius having the
greatest number of arcs is $r_m$-critical. By Theorem \ref{th4},
the Hertz graph of the graph $D$ is isomorphic to the graph
$\Gamma_{k,s;k_1,...,k_s}$ for some $s,k_1,...,k_s$. Let
$\{Y_1,...,Y_s\}$ be the decomposition of the vertex set of the graph
$\Gamma_{k,s;k_1,...,k_s}$ described above. The graph induced by the
set $Y_\alpha$ is isomorphic to $\Gamma_{k_\alpha,0}$. Let $D_\alpha$ be
the subgraph of the graph $D$ induced by bicomponents corresponding to
vertices of the set $Y_\alpha$. It is easy to see that all bicomponents
of the graph $D_\alpha$, excluding perhaps one not corresponding to the
vertex 1 of the graph $\Gamma_{k_\alpha,0}$, are one-vertex. If it is
not so, then we replace the subgraph $D_\alpha$ of $D$ by a graph
$D_\alpha'$
possessing
the above property. Then evidently, the number of
arcs connecting vertices of the subgraph $D_\alpha$ to other vertices of
the graph $D$ equals  the number of arcs connecting vertices of the
subgraph $D_\alpha'$ to other vertices of the graph $D$; and the number
of arcs in the graph $D\alpha'$ is greater than the number of arcs in
the graph $D_\alpha$. Suppose that bicomponents corresponding to the
vertices $k_\alpha$ of the graphs $\Gamma_{k_\alpha,0}$  are the only
bicomponents that can have more than one vertex. Let $|Y_i|\ge 4$.
Consider then a new decomposition
$\{Y_1,...,Y_{i-1},Y_i',Y_i'',Y_{i+1},...,Y_s\}$ of the vertex set of
the graph $\Gamma_{k,s;k_1,...,k_s}$. Here $Y_i'$ consists of the two
vertices of the set $Y_i$ corresponding to the vertices $2,3$ of the
graph $\Gamma_{k_i,0}$, and $Y_i''$ consists of the rest vertices of the
set $Y_i$. The difference between the graphs $\Gamma_{k,s;k_1,...,k_s}$
and $\Gamma_{k,s+1;k_1,...,k_{i-1},2,k_i-2,k_{i+1},...,k_s}$ is that the
arc $(2,3)$ is missing in the subgraph induced by the set $Y_i$ in the
graph $\Gamma_{k,s;k_1,...,k_s}$; but the
arcs $(2,1)$ and $(3,1)$ that are missing in the graph
$\Gamma_{k,s;k_1,...,k_s}$ are present in
$\Gamma_{k,s+1;k_1,...,k_{i-1},2,k_i-2,k_{i+1},...,k_s}$.
Taking into account that bicomponents of the graph $D$ corresponding to
the vertices $1,2,3$ of the graph $\Gamma_{k_i,0}$ are one-vertex, we
see that the number of arcs in the graph $D$ increases by one after the
above operation. Thus, we have shown that $|Y_\alpha|\le 3$
$(\alpha=1,...,s)$ in the Hertz graph of the graph $D$. Suppose now that
the Hertz graph $\Gamma$ of the graph $D$ contains two subsets $Y_i$ and
$Y_j$ such that $|Y_i|=|Y_j|=3$. Let us consider the sets $\{2^i,2^j\}$,
$\{1^i,3^i\}$, $\{1^j,3^j\}$ instead of the sets $Y_i$ and $Y_j$
where $\alpha^\beta$ is the vertex of the
graph $\Gamma$ corresponding to the vertex $\alpha$ of the graph
$\Gamma_{k_\beta,0}$. Here if $i<j$ then the arcs $(2^i,2^j)$,
$(1^i,2^j)$, $(3^i,2^j)$ vanish from the graph $\Gamma$, and the arcs
$(2^i,1^i)$, $(2^j,1^i)$, $(2^j,3^i)$, $(2^i,1^j)$ appear in it.
Note
that the bicomponents of the graph $D$ corresponding to the
vertices $1^i,2^i,1^j,2^j$ are one-vertex.
Hence
the described operation results in increasing the number of arcs
of the graph $D$ by one.

The above considerations imply that if an $n$-vertex graph $D$ of
infinite quasi-radius with $k$ bicomponents contains the greatest number
of arcs, then its Hertz graph is isomorphic to the graph
$\Gamma_{k,s;k_1,...,k_s}$ where $s=\intt{\sfrac{k}{2}}$. It is easy
to see that the number of arcs in the $n$-vertex digraph $D$ whose Hertz
graph is isomorphic to $\Gamma_{k,s;k_1,...,k_s}$,
$s=\intt{\sfrac{k}{2}}$ is maximal, if all bicomponents of the graph
$D$, excluding perhaps one not corresponding to the vertex 1 of the
subgraph of $\Gamma_{k,s;k_1,...,k_s}$ induced by the set $Y_i$,
$|Y_i|=3$, are one-vertex. In this case a simple calculation shows that
the graph $D$ contains $n(n-k-1)+\intt{\sfrac{k^2}{2}}$ arcs.
\rule{2mm}{2mm}

\begin{Cor} \label{cor42}
If a digraph $G$ with $k$ bicomponents and infinite quasi-radius has the
property that the addition of an arbitrary missing arc converts it into
a digraph with finite quasi-radius, then

a) $k=2l$;

b) the Hertz graph of $G$ is isomorphic to $\Gamma_{k,k/2;2,...,2}$.
\end{Cor}

{\bf Proof.\/}
Suppose that the Hertz graph of the digraph $G$ is isomorphic to
$\Gamma_{k,s;k_1,...,k_s}$ and $\exists i\,(k_i\ge 3)$. As we have
stated, the subgraph induced by the set $Y_i$ is isomorphic to
$\Gamma_{k_i,0}$. It is easy to see that if we add the arc $(3^i,2^i)$
to the Hertz graph of the graph $G$, then the quasi-radius of the
obtained graph is infinite. Hence $\forall i\, (k_i=2)$ and $k=2l$.
\rule{2mm}{2mm}

\bigskip

Let $\mypi(n,k)$ denote the number of nonisomorphic $n$-vertex
$r_m$-critical digraphs with $k$ bicomponents and infinite
quasi-radius.

\begin{Cor} \label{cor43}
\begin{eqnarray*}
&&\hspace{-\parindent}
\mypi(n,k)=\!
\dsum_{l=1}^{\intt{k/2}}\dsum_{s=(3l-k)_+}^{l}
\dsum_{t=2s}^{n-k+2s}\dsum_{p_1+...+p_s=t}
{\biggl(
\raisebox{-0.9ex}{\mbox{$\stackrel {\textstyle {l}}{\mathstrut s}$}}
\biggr)}
\binom{k-2l-1}{l-s-1}
\binom{n-t-1}{k-2s-1}
\left\{
\prod_{i=1}^{s}\intt{\frac{p_i}{2}}
\right\} ,\\
&&\hspace{-\parindent}
\mbox{where }m_+=\max\{m,0\},\quad
\binom{0}{k}\stackrel{\rm df}{=}
\left\{
\begin{array}{l}
0, k\ne 0,\\
1, k=0,
\end{array}
\right.\
\binom{-1}{-1}\stackrel{\rm df}{=}1,\quad
{\rm and}\
\prod_{i=1}^{0}\intt{\frac{p_i}{2}}\stackrel{\rm df}{=}1.
\end{eqnarray*}
\end{Cor}

{\bf Proof.\/}
First we note that if the Hertz graphs of digraphs $G_1$ and $G_2$ are
nonisomorphic, then the digraph $G_1$ and $G_2$ are not isomorphic
either. Let us count the number of nonisomorphic $n$-vertex $r_m$-critical
digraphs with $k$ bicomponents and $r_m=\infty$ such that their
Hertz graphs have $l$ blocks $Y_1,...,Y_l$, exactly $s$ of which are
two-vertex. This number equals the number of ways of
distributing
$k$ identical particles in $l$ distinct boxes such that each box contains
at least two particles and exactly $s$ boxes contain two particles each.
The latter number equals (see \cite{A79})
$$
{\biggl(
\raisebox{-0.9ex}{\mbox{$\stackrel {\textstyle {l}}{\mathstrut s}$}}
\biggr)}
\binom{k-2l-1}{l-s-1}.
$$
We note that distinct
distributions
(by number) of vertices in
nontwo-vertex blocks correspond to nonisomorphic graphs.

Let us consider some two-vertex block $Y_i$ that contains $p_i$
vertices. Suppose the position of vertices in all other blocks is fixed.
Then varying the number of vertices only in the vertices of the block
$Y_i$ we can obtain $\intt{\sfrac{p_i}{2}}$ nonisomorphic graphs. Let
the number of vertices of the digraph contained in the two-vertex blocks
equal $p_1+...+p_s=t$. The number of ways of
distributing
the rest $n-t$
vertices in the vertices of nontwo-vertex blocks equals
$\binom{n-t-1}{k-2s-1}$; and the number of ways of
distributing
$t$ vertices in the vertices of $s$ two-vertex blocks (such that the
obtained graphs are nonisomorphic) equals
$\dsum_{p_1+...+p_s=t}\left\{\prod_{i=1}^{s}\intt{\frac{p_i}{2}}\right\}$
Finally we obtain that the number of nonisomorphic $n$-vertex
$r_m$-critical digraphs with $k$ bicomponents and $r_m=\infty$,
the Hertz graphs of which have $l$ blocks exactly $s$ of which are
two-vertex, equal
\begin{eqnarray*}
&&\hspace{-\parindent}
\dsum_{p_1+...+p_s=t}
{\biggl(
\raisebox{-0.9ex}{\mbox{$\stackrel {\textstyle {l}}{\mathstrut s}$}}
\biggr)}
\binom{k-2l-1}{l-s-1}
\binom{n-t-1}{k-2s-1}
\left\{
\prod_{i=1}^{s}\intt{\frac{p_i}{2}}
\right\}
\end{eqnarray*}
Summing the quantity over $t,s,l$ we obtain what we needed to prove.
\rule{2mm}{2mm}

\bigskip

Let $\xi(n,k)$ denote the number of distinct $r_m$-critical
digraphs with $k$ bicomponents and infinite quasi-radius that can be
constructed on $n$ numbered vertices.

\begin{Cor} \label{cor44}
\begin{eqnarray*}
&&\hspace{-\parindent}
\xi(n,k)=\!
\dsum_{l=1}^{\intt{k/2}}\dsum_{s=(3l-k)_+}^{l}
\dsum_{t=2s}^{n-k+2s}\dsum_{p_1+...+p_s=t}
{\biggl(
\raisebox{-0.9ex}{\mbox{$\stackrel {\textstyle {l}}{\mathstrut s}$}}
\biggr)}
\binom{k-2l-1}{l-s-1}
\binom{n}{n-t}\times\\
&& \times
(k-2s)!\, S(n-t,k-2s)\frac{t!}{p_1!...p_s!}
\left\{
\prod_{i=1}^{s}(2^{p_i-1}-1)
\right\} ,
\end{eqnarray*}
where $S(u,v)$ are the Stirling numbers of the second kind.
\end{Cor}

{\bf Proof.\/}
The proof of the corollary is analogous to the proof of Corollary
\ref{cor43}; we just note that
$(k-2s)!\, S(n-t,k-2s)$ is the number of ways of
distributing
$n-t$ numbered vertices in $k-2s$ bicomponents contained in nontwo-vertex
blocks;\\
$\dsum_{p_1+...+p_s=t} \frac{t!}{p_1!...p_s!}
\left\{ \prod_{i=1}^{s}(2^{p_i-1}-1) \right\}$ is the number of ways of
distributing
$t$ numbered vertices in $2s$ bicomponents contained in
two-vertex blocks, and there are $\binom{n}{n-t}$ ways to choose $t$
vertices contained in bicomponents of two-vertex blocks.
\rule{2mm}{2mm}


\bigskip
\begin{center}
\large \bf
3.
 On maximal digraphs of finite \\
   radius and quasi-diameter
\end{center}

In the preceding sections we have characterized up to isomorphism critical
digraphs with infinite values of $d,d_m,r,r_m$. $d$-critical
digraphs of finite diameter were characterized up to isomorphism by
L.S.~Mel'nikov \cite{M70}. The solution to the corresponding problem for
critical digraphs with finite values of $d_m,r,r_m$ has not been found
yet. However, we succeeded in obtaining the least upper bounds on the
number of arcs in $n$-vertex digraphs with given finite values of radius
and quasi-diameter, and characterizing the digraphs for which the bounds are
achieved.
We should remark in this connection that the least upper
bound on the number of arcs in an $n$-vertex digraph of given finite
radius was found by \u{S}.M.~Ismailov \cite{I73}. However, there are
significant ambiguities in the proof of the result in \cite{I73}
and it seems to us that the method used by author can't lead to the
proof.
The least upper bound on the number of edges in an ordinary undirected
$n$-vertex graph of given radius was obtained by V.G.~Vizing \cite{V67}.

\bigskip

{\bf I. }Let $G$ be an $n$-vertex directed graph without loops and
$r(G)=k<\infty$. We remind that the digraph $G$ is said to be
{\em maximal\/} if it has the maximum number of arcs among all
$n$-vertex digraphs of radius $k$.

In this subsection we shall obtain the least upper bound on the number
of arcs in an $n$-vertex digraph of radius $k$, and
characterize all maximal
digraphs.

Let a digraph $G=(X,U)$ have radius $k$. A vertex $x_1$ of the graph $G$
is called a {\em center\/}, if $\max\limits_{y\in X}\rho(x_1,y)=r(G)=k$.
There exists a vertex $x_{k+1}$ in $G$ such that $\rho(x_1,x_{k+1})=k$.
Let $\{x_1,...,x_{k+1}\}$ be a naturally ordered totality of vertices of
some shortest directed path going from $x_1$ to $x_{k+1}$. Let vertices
$x_1,...,x_t$ belong to one bicomponent, and vertices
$x_{t+1},...,x_{k+1}$ do not belong to this bicomponent. Denote the set
$\{x_1,...,x_{k+1}\}$ by $M$. Partition the set $X\setminus M$ into two
subsets $S$ and $Y$ in the following way: the set $S$ consists of the
vertices that belong to the bicomponent of $G$ containing the vertex
$x_1$, and the set $Y$ consists of all other vertices of the set
$X\setminus M$. Let the set $S$ contain $s$ vertices.

Denote by $g(n,k)$ the maximum number of arcs in a directed $n$-vertex
graph of radius $k$. It is obvious that $g(n,1)=n(n-1)$, and
$g(n,2)=n(n-2)$; and a digraph of radius two is maximal iff the outdegree
of each vertex  of the graph equals $n-2$. Thus, it remains
to consider the case $k\ge 3$.

Let us define a function $F(n,k,s,t)$ in the following way:
\begin{eqnarray*}
F(n,k,s,t)&=&n(n-k)+\frac{k^2-k-2}{2}+\\
&& +\left[ -n(s+t+2)+s^2+ts+t^2+3k+2+\right.\\
&& +\,\Bigl.(n-k-s-1)\max\{t,3\}+s\max\{k,t+2\}\Bigr].
\end{eqnarray*}


\begin{Lem} \label{lem8}
If $1\le t\le k$ then the number of arcs in an $n$-vertex digraph of
radius $k$ does not exceed
$\max\limits_{1\le t\le k}\max\limits_{0\le s\le n-k-1}F(n,k,s,t)$.
\end{Lem}


{\bf Proof.\/}
Let us consider Figure 6.

\begin{figure}[hbpt]
\begin{picture}(150,55)

\put(45,50){\line(1,0){15}}               %
\put(90,50){\line(1,0){15}}               %
\put(45,50){\circle*{1}}                  %
\put(45,53){\makebox(0,0){$x_1$}}         %
\put(60,50){\circle*{1}}                  %
\put(60,53){\makebox(0,0){$x_2$}}         %
\put(75,50){\circle*{1}}                  %
\put(75,53){\makebox(0,0){$x_t$}}         %
\put(90,50){\circle*{1}}                  %
\put(90,53){\makebox(0,0){$x_k$}}         %
\put(105,50){\circle*{1}}
\put(105,53){\makebox(0,0){$x_{k+1}$}}    %
\put(67.5,50){\makebox(0,0){\ldots}}      %
\put(82.5,50){\makebox(0,0){\ldots}}      %
\put(60,30){\circle{13}}                  %
\put(60,30){\makebox(0,0){\LARGE $S$}}    %
\put(90,30){\circle{13}}                  %
\put(90.5,30){\makebox(0,0){\LARGE $Y$}}  %
\put(75,15){\makebox(0,0){Fig. 6}}      %

\thicklines
\put(54,50){\vector(1,0){0}}                  
\put(99,50){\vector(1,0){0}}                  
\end{picture}
\end{figure}

\vspace{-15mm}
{\em (i)\/} Let us bound the number of arcs connecting the sets $M$ and
$Y$. Since all vertices of the set $Y$ do not belong to the bicomponent
containing the vertex $x_1$, no arc goes from a vertex of the set $Y$ to
a vertex of the set $\{x_1,...,x_t\}$. Let $y\in Y$ and the vertex $y$
be adjacent to some vertex $x_l \in M$. Then, a fortiori, no arc goes
from the vertex $y$ to the vertices of $M$ whose numbers are greater
than $l+2$. Taking this into account, one obtains easily that the number
of arcs connecting the vertex $y$ to vertices of the set $M$ does not
exceed $(k+1-t)+\max\{t,3\}$; hence the number of arcs connecting  the
sets $M$ and $Y$ does not exceed $(n-s-k-1)(k+1-t+\max\{t,3\})$. In
addition, the number of arcs in the subgraph induced by the set $Y$
does not exceed $(n-s-k-1) (n-s-k-2)$.

{\em (ii)\/} Let us bound the number of arcs incident to the set $S$.
Let $z$ be an arbitrary vertex in the set $S$.

a) Suppose that no arc goes from vertices of the set $M$ to the vertex
$z$. Here if $k+1$ arcs go from the vertex $z$ to the set $M$, then
there is a vertex in the set $(S\cup Y)\setminus \{z\}$ that is not a
terminal vertex of an arc issuing from $z$; since otherwise, the vertex
$z$ would be a center in $G$, and the radius of $G$ would equal 1,
contrary to the hypothesis. Hence in this case the number of arcs going
from the vertex $z$ to other vertices of $G$ does not exceed
$k+(s-1)+(n-k-s-1)=n-2$.

b) Suppose there are arcs in $G$ that go from the set $M$ to the vertex
$z$. We note that the vertices $x_1,...,x_t$ are the only vertices with
this property. Moreover, if an arc goes from a vertex $x_l$ to the vertex
$z$, then no arc can go from the vertex $z$ to the vertex $x_i$ for
$i>l+2$; since otherwise, $\rho(x_1,x_{k+1})$ would be less than $k$.
So it is easy to obtain that the
number of arcs connecting the vertex $z$ to the vertices of $M$ does not
exceed $t+3$. If this number equals $t+3$ then arcs go from the vertex
$z$ to the vertices
$x_1,x_2,x_3,\, \max\limits_{1\le i\le k+1}\rho(z,x_i)<k$;
and in order that radius of the graph $G$ be greater or equal to $k$ it
is necessary that at least one arc of the kind $(z,u)$ where
$u\in (S\cup Y)\setminus\{z\}$ is missing in $G$. Then the total number
of arcs connecting the vertex $z$ to the set $M$ and going from $z$ to
the set $S\cup Y$ does not exceed $(t+2)+(s-1)+(n-s-k-1)=n+t-k$; and the
number of arcs incident to the set $S$ does not exceed $s(n+t-k)$ in this
case. Combining cases a) and b) we see that the number of arcs incident
to the set $S$ does not exceed $s\max\{n-2,n+t-k\}$.

{\em (iii)\/} Let us now bound the number of arcs in the subgraph
induced by the set $M$.
Since $t\le k$,
hence no arc goes from a vertex whose number is greater than $t$ to a vertex
whose number is less or equal to $t$. Hence the number of arcs in this
subgraph does not exceed
$$
k+(t-1)+(t-2)+...+1+(k-t)+(k-t-1)+...+1=
k+\sfrac{t(t-1)}{2}+\sfrac{(k-t+1)(k-t)}{2}.
$$

It is easy to see that in these three cases we have considered all arcs
in the graph $G$. Hence if $t\le k$ and $s$ is fixed
$(0\le s\le n-k-1)$, then the number of arcs in $G$ does not exceed
\begin{eqnarray*}
&& (n-s-k-1)(k+1-t+\max\{t,3\})+(n-s-k-1)(n-s-k-2)+\\
&& +k+t^2-t-kt+\frac{k^2+k}{2}+s\max\{n-2,n+t-k\}=\\
&& =n(n\!-\!k)+\frac{k^2\!-\!k\!-\!2}{2}+
\Bigl[(n\!-\!s\!-\!k\!-\!1)\max\{t,3\}+s\max\{n\!-\!2,n+t\!-\!k\}+\Bigl.\\
&& \left. +3k+2+s^2+st+sk+2s-2sn+t^2-tn-2n\right]=F(n,k,s,t).
\ \ \ \rule{2mm}{2mm}
\end{eqnarray*}

\begin{Lem} \label{lem9}
For $1\le t\le k-1$
$$
F(n,k,s,t)\le n(n-k)+\sfrac{k^2-k-2}{2},
$$
and an equality holds iff $t=1$ and $s=0$.
\end{Lem}

{\bf Proof.\/}
a) Suppose $t<3,\ k\ge t+2$. Then
\begin{eqnarray*}
&& F(n,k,s,t)= n(n-k)+\frac{k^2-k-2}{2}+3n-3s-3k-3+sn-2s+\\
&& +3k+2+s^2+sk+st+2s-2sn+t^2-tn-2n=\\
&& =n(n-k)+\frac{k^2-k-2}{2}-s(n-s-k-t+3)-(t-1)(n-t-1)\le\\
&& \le n(n-k)+\frac{k^2-k-2}{2}.
\end{eqnarray*}
Here an equality in the last inequality holds iff $t=1$ and $s=0$.

\medskip
b) Suppose $t\ge 3,\ k\ge t+2$. Then
\begin{eqnarray*}
&& F(n,k,s,t)= ns-st-kt-t+sn-2s+3k+2+s^2+sk+\\
&& +st+2s-2sn+t^2-tn-2n+n(n-k)+\frac{k^2-k-2}{2}=\\
&& =n(n-k)+\frac{k^2-k-2}{2}-s(n-k-s)-2(n-k-1)-(t-1)(k-t)<\\
&& <n(n-k)+\frac{k^2-k-2}{2}.
\end{eqnarray*}

c) Suppose $t<3,\ k<t+2$. Then $t=2,k=3$ and hence
\begin{eqnarray*}
F(n,3,s,2)&=&n(n-3)+\frac{3^2-3-2}{2}-n+3-s(n-s-3)<\\
&& <n(n-3)+\frac{3^2-3-2}{2}=n^2-3n+2.
\end{eqnarray*}

d) Suppose $t\ge 3,\ k<t+2$. Since $t\le k-1$ in order to prove the
lemma
it suffices to consider the case $t=k-1$. Then
\begin{eqnarray*}
&& F(n,k,s,k-1)= n(n-k)+\frac{k^2-k-2}{2}+
\left[-n(s+k+1)+s^2+(k-1)s+\right.\\
&& \left.+(k-1)^2+3k+2+(n-k-s-1)(k-1)+s(k+1)\right]=\\
&& =n(n-k)+\frac{k^2-k-2}{2}-s(n-s-k-1)-2n+k-4<\\
&& <n(n-k)+\frac{k^2-k-2}{2}. \ \ \ \rule{2mm}{2mm}
\end{eqnarray*}

\begin{Lem} \label{lem10}
Let $G$ be an $n$-vertex digraph with radius $k<\infty$, $A_1$ be the
bicomponent that contains a center. Then the outdegree of any
vertex in $A_1$ does not exceed $n-k$.
\end{Lem}

{\bf Proof.\/}
Let $v$ be an arbitrary vertex in the bicomponent $A_1$, and $B_v$ be
the totality of those vertices in $G$ to which arcs go from $v$.
$|B_v|=m$ is the outdegree of the vertex $v$. We note that any
vertex in the graph $G$ is reachable from the vertex $v$, and the
shortest path from $v$ to an arbitrary vertex in $G$ contains at most
one vertex of the set $B_v$. Hence the path contains at most $n-m+1$
vertices, and its length does not exceed $n-m$. It means $r(G)\le n-m$;
since $r(G)=k$ we have $m\le n-k$.
\rule{2mm}{2mm}

\bigskip

{\bf Remark.} It follows from Lemma \ref{lem10} that the number of arcs
in an $n$-vertex biconnected digraph of radius $k$ does not exceed
$n(n-k)$.

\begin{Th} \label{th5}
The following equalities hold:
\begin{eqnarray*}
&& g(n,1)=n(n-1),\quad g(n,2)=n(n-2),\\
&& g(n,k)=n(n-k)+\frac{k^2-k-2}{2}\mbox{  for }k\ge 3.
\end{eqnarray*}
\end{Th}

{\bf Proof.\/}
We only need to prove the third relation.

Denote the quantity $n(n-k)+\sfrac{k^2-k-2}{2}$ by $\varphi(n,k)$. First we
shall show that $g(n,k)\le\varphi(n,k)$. It follows from Lemmas
\ref{lem8},\,\ref{lem9} that if $t<k$ in an $n$-vertex digraph of radius
$k$, then this inequality holds. Partition the vertex set of the graph
$G$ into two subsets $A_1$ and $A_2$, where $A_1$ is the bicomponent
containing center, and $A_2$ consists of all other vertices. If $t\ge k$
then $|A_1|=p\ge k$. The number of arcs in the subgraph induced by the
set $A_2$ does not exceed $(n-p)(n-p-1)$. Now we note that if we sum
outdegrees of all vertices in the set $A_1$, then we obtain the
number of all other arcs in the graph $G$. But by Lemma \ref{lem10}, the
outdegree of any vertex in the set $A_1$ does not exceed $n-k$.
Hence the number of arcs in the graph $G$ does not exceed
$p(n-k)+(n-p)(n-p-1)=n(n-k)-(p-k+1)(n-p)$. Taking into account that in
our case $n\ge p\ge k$ we obtain that the latter expression does not
exceed $n(n-k)$. The inequality $g(n,k)\le\varphi(n,k)$ is proved.

In order to prove the reverse inequality we consider the following
$n$-vertex digraph $D_0$: The vertex set of the digraph is partitioned
into $k+1$ disjoint subsets $X_1,...,X_{k+1}$ where
$|X_1|=|X_3|=...=|X_{k+1}|=1$ and $|X_2|=n-k$. An arc goes from each
vertex in $X_i$ to each vertex in $X_{i+1}$, $i=1,...,k$;
if $1<i<j\le k+1$ then an arc goes from each vertex in $X_j$ to each
vertex in $X_i$. The subgraph induced by the set $X_2$ is complete
symmetric. There are no other arcs in the digraph $D_0$. Evidently, the
radius of the graph equals $k$. A simple calculation shows that the
number of arcs in this graph equals $\varphi(n,k)$. This proves the
inequality $g(n,k)\ge\varphi(n,k)$.
\rule{2mm}{2mm}

\bigskip

Now we take up
characterizing maximal digraphs of given radius. It follows
from the proof of Theorem \ref{th5} that in a maximal digraph
$S=\emptyset, t=1$, the subgraph induced by the set $Y$ is complete
symmetric, and the subgraph induced by the set $M$ has the following
properties: an arc goes from the vertex $x_i$ to the vertex $x_{i+1}$;
if $1<i<j\le k+1$ then an arc goes from the vertex $x_j$ to the vertex
$x_i$; there are no other arcs in the subgraph. Let $x_i$ be the first
vertex from which arcs go to the set $Y$, and $Y_i$ be the totality of
vertices in $Y$ to which arcs go from $x_i$. Thus, if $i=1$ then each
vertex of the set $Y_1$ is connected by a pair of antiparallel arcs to
the vertices $x_2$ and $x_3$, and an arc goes from each vertex of the
set $\{x_4,...,x_{k+1}\}$ to each vertex of the set $Y_1$. Each vertex
in the set $Y\setminus Y_1$ is connected by a pair of antiparallel arcs
to each of the vertices $x_2,x_3,x_4$, and an arc goes from each vertex
in the set $\{x_5,...,x_{k+1}\}$ to each vertex in the set
$Y\setminus Y_1$. Partition the vertex set of this graph into $k+1$
subsets $X_1,...,X_{k+1}$ as follows:
$$
X_1=\{x_1\},\ X_2=\{x_2\}\cup Y_1,\ X_3=\{x_3\}\cup(Y\setminus Y_1),\
X_4=\{x_4\},...,\ X_{k+1}=\{x_{k+1}\}.
$$

We see that the subgraphs induced by each of these subsets are complete
symmetric; an arc goes from each vertex in the set $X_i$ to each vertex
in the set $X_{i+1}$; if $1<i<j\le k+1$ then an arc goes from each
vertex in the set $X_j$ to each vertex in the set $X_i$; there are no
other arcs in this graph.

Let $x_i$ be the first vertex in the set $M$ from which arcs go to the
set $Y$, and $i>1$. Then each vertex in the set $Y_i$ has the following
properties: an arc goes from each vertex in the set $Y_i$ to each vertex
in the set $\{x_2,...,x_{i-1}\}$; each vertex in the set $Y_i$ is
connected by a pair of antiparallel arcs to the vertices $x_i,x_{i+1},
x_{i+2}$; and an arc goes from each vertex in the set
$\{x_{i+3},...,x_{k+1}\}$ to each vertex in the set $Y_i$. Each vertex
in the set $Y\setminus Y_i$ has the following properties: an arc goes
from each vertex in the set $Y\setminus Y_i$ to each vertex  in the set
$\{x_2,...,x_i\}$, each vertex in the set $Y\setminus Y_i$ is connected
by a pair of antiparallel arcs to each vertex in the set
$\{x_{i+1},x_{i+2},x_{i+3}\}$; an arc goes from each vertex in the set
$\{x_{i+4},...,x_{k+1}\}$ to each vertex in the set $Y\setminus Y_i$.
Let us partition the vertex set of this graph into nonempty disjoint
subsets $X_1,...,X_{k+1}$ as follows:
\begin{eqnarray*}
&& X_1=\{x_1\},...,X_i=\{x_i\},\ X_{i+1}=\{x_{i+1}\}\cup Y_i,\\
&& X_{i+2}=\{x_{i+2}\}\cup(Y\setminus Y_1),\ X_{i+3}=\{x_{i+3}\},...,
X_{k+1}=\{x_{k+1}\}.
\end{eqnarray*}
The decomposition
possess
the same properties as those of the
decomposition in the preceding case. Thus we have proved

\begin{Th} \label{th6}
All $n$-vertex maximal digraphs $D$ of finite radius $k$ are exhausted
by the following:

a) if $k=1$, then $D$ is a complete symmetric digraph;

b) if $k=2$, then $D$ is a digraph such that the outgoing semidegree of
each vertex equal $n-2$;

c) if $k\ge 3$, then the vertex set of $D$ can be partitioned into $k+1$
nonempty disjoint subsets $X_1,...,X_{k+1}$ such that
$|X_1|=|X_{k+1}|=1$, all other subsets, excluding perhaps two with
consecutive indices, are singletons; if $1<i<j\le k$, then an arc goes
from each vertex of the set $X_j$ to each vertex of the set $X_i$; an
arc goes from each vertex of the set $X_i$ to each vertex of the set
$X_{i+1}$, $i=1,...,k$; the subgraphs induced by each of the subsets
$X_i$ are complete symmetric; there are no other arcs in this graph
besides those listed above.
\end{Th}

\begin{Cor} \label{cor51}
For $k\ge 3$ the number of nonisomorphic $n$-vertex maximal digraph of
radius $k$ equals $(n-k-1)(k-2)+1$.
\end{Cor}

{\bf Proof.\/}
It follows from Theorem \ref{th6} that the number of such graphs equals
the number of ways of
distributing
$n-2$ identical particles in $k-1$
nonnumbered boxes without empty ones such that each box, excluding
perhaps two neighboring ones, contains one particle. For $n=k+1$ the
number equals 1; for $n=k+1$ it equals $k-1$. For $n\ge k+3$ the number
of ways such that there is a box containing $n-k-1$ particles equals
$k-1$; and the number of ways such that exactly two neighboring boxes
contain more than one particle equals $(n-k-2)(k-2)$. Hence the total
number equals $(n-k-2)(k-2)+k-1=(n-k-1)(n-2)+1$. The last formula covers
all the three cases.
\rule{2mm}{2mm}

\begin{Cor} \label{cor52}
Let $\chi(n,k)$ denote the number of maximal $n$-vertex graphs of radius
$k$ that can be constructed on given $n$ numbered vertices. Then
$$
\chi(n,k)=\left\{
\begin{array}{ll}
1 & \mbox{for }k=1,\\
(n-1)^n & \mbox{for }k=2,\\
(k\!-\!1)k!\binom{n}{k}+
(k\!-\!2)(k\!-\!1)!\binom{n}{k\!-\!1}(2^{n-k+1}\!-\!2n+2k\!-\!4)
& \mbox{for }k\ge 3.
\end{array}
\right.
$$
\end{Cor}

The proof of the corollary is analogous to the proof of Corollary
\ref{cor51}.

\bigskip

{\bf II. }In this subsection we shall obtain the least upper bound on
the number of arcs in an $n$-vertex digraph of quasi-diameter
$k<\infty$, and
characterize maximal digraphs of finite quasi-diameter.

We note that the only maximal digraph of quasi-diameter 1 is a complete
symmetric graph; a digraph is maximal of quasi-diameter 2 iff it is
isomorphic to a complete symmetric graph with a pair of arcs of the kind
$(x,y),(y,x)$ removed. Thus it only remains to consider the case $k\ge 3$.

Let $G=(X,U)$ be an $n$-vertex digraph and $d_m(G)=k$. Hence there
exists a pair of vertices $x_1,x_{k+1}$ in $G$ such that
$\rho_m(x_1,x_{k+1})=\min\{\rho(x_1,x_{k+1}),\rho(x_{k+1},x_1)\}=k$.
Let for instance, $\rho(x_1,x_{k+1})=k$, and
$\{x_1,x_2,...,x_k,x_{k+1}\}$ be the naturally ordered totality of
vertices of some shortest directed path from $x_1$ to $x_{k+1}$. Denote
the set $\{x_1,...,x_{k+1}\}$ by $M$, and the set $X\setminus M$ by $B$.
Evidently, the quasi-diameter of the subgraph induced by the set $M$
equals $k$. Let us bound the number of arcs in the subgraph induced by
the set $M$.

\begin{Lem} \label{lem11}
A $(k+1)$-vertex digraph $D$ of quasi-diameter $k$ contains at most
$\sfrac{k^2+k}{2}$ arcs.
\end{Lem}

{\bf Proof.\/}
{\em (i)\/} Suppose that the graph $D$ is not biconnected; let vertices
$x_1,...,x_p$ be all vertices of one of its bicomponents. It is easy to
see that the number of arcs in the graph $D$ does not exceed
\begin{eqnarray*}
&& k+(p-1)+(p-2)+...+1+(k-p)+(k-p-1)+...+1=\\
&& =k+\frac{p^2-p}{2}+\frac{(k-p+1)(k-p)}{2}\le\frac{k^2+k}{2},
\end{eqnarray*}
Here the equality holds iff $p=1$ or $p=k$.

{\em (ii)\/} Suppose that the graph $D$ is biconnected. Then it is easy
to see that both outdegree and indegree of the
vertices $x_1$ and $x_{k+1}$ do not exceed 1. Hence the number of arcs
in the digraph $D$ does not exceed
$$
(k-2)+\left[(k-2)+(k-3)+...+1\right]+4=\sfrac{k^2+k}{2}-(k-3)\le
\sfrac{k^2+k}{2},
$$
and the equality holds iff $k=3$.
\rule{2mm}{2mm}

\bigskip

Now let us bound the number of arcs that connect the sets $M$ and $B$.
It is easy to see that each vertex in the set $B$ can be connected by
a pair of
antiparallel arcs to at most three vertices in the set $M$. Here if a
vertex $z\in B$ is connected by a pair of antiparallel arcs to three
vertices in the set $M$, then it is easy to see that the vertex $z$ is
not adjacent either to the vertex $x_1$ or to the vertex $x_{k+1}$.
Hence the number of arcs connecting an arbitrary vertex $z\in B$ to the
set $M$ does not exceed $k+3$; the number of arcs connecting the sets
$M$ and $B$ does not exceed $(k+3)(n-k-1)$; and the number of arcs in
the subgraph induced by the set $B$ does not exceed $(n-k-1)(n-k-2)$.
Thus, the number of arcs in an $n$-vertex graph of quasi-diameter $k$
does not exceed
\begin{eqnarray*}
&& \frac{k^2+k}{2}+(k+3)(n-k-1)+(n-k-1)(n-k-2)=\\
&& =\frac{k^2+k}{2}+(n-k-1)(n+1)=n(n-k)+\frac{k^2-k-2}{2}
\end{eqnarray*}

Denote by $f(n,k)$ the
maximum
number of arcs in an $n$-vertex digraph
of quasi-diameter $k$.

\begin{Th} \label{th7}
\begin{eqnarray*}
&& f(n,1)=n(n-1),\quad f(n,2)=n(n-1)-2,\\
&& f(n,k)=n(n-k)+\frac{k^2-k-2}{2}\mbox{ for }k\ge 3.
\end{eqnarray*}
\end{Th}

{\bf Proof.\/}
We only need to prove the last relation, moreover, the inequality
$f(n,k)\le n(n-k)+\sfrac{k^2-k-2}{2}$ has already been proved. In order
to prove the reverse inequality we note that for $k\ge 3$ the
quasi-diameter of any maximal $n$-vertex digraph of radius $k$ equals
$k$. But by Theorem \ref{th5}, the number of arcs in a maximal $n$-vertex
digraph of radius $k$ equals $n(n-k)+\sfrac{k^2-k-2}{2}$.
\rule{2mm}{2mm}

\bigskip

Now we proceed to
characterizing maximal digraphs of given quasi-diameter.
Suppose $k\ge 4$. Then it follows from Lemma \ref{lem11} that the
subgraph induced by the set $M$ is not biconnected, and $p=1$ or
$p=k$. If $p=1$ then it is easy to see that maximal digraphs
of quasi-diameter $k$ coincide with maximal digraphs of radius $k$
described
in Theorem \ref{th6}. Suppose now $p=k$. We note that in view
of the symmetry of the function $\rho_m(x,y)$ a graph $D$ is maximal of
quasi-diameter $k$ iff
so is the graph $\mybarr{D}$ that is obtained from the graph $D$ by
reversing the direction of its arcs. But if we reverse the direction of
arcs in the graph $D$ in which $p=k$, then we obtain the graph
$\mybarr{D}$ in which $p=1$. Thus, we have shown that for $k\ge 4$ a
maximal digraph of quasi-diameter $k$ is either a maximal digraph of
radius $k$ or is obtained from a maximal digraph of radius $k$ by
reversing the direction of its arcs.

Let us now consider the case $k=3$. We note that if a maximal digraph of
quasi-diameter 3 is not biconnected, then all the above considerations
are applicable to it, and hence it is either a maximal digraph of radius
3 or is obtained from a maximal digraph of radius 3 by reversing the
direction of its arcs.

Let a maximal digraph $G=(X,U)$ of quasi-diameter 3 be biconnected. Then
it is easy to verify that there exist vertices $z,y,u,v$ in the graph
such that $\rho_m(z,v)=\rho(z,v)$ and the subgraph induced by the set
$\{z,y,u,v\}$ is biconnected. Then the subgraph is isomorphic to either
of the graphs in Figures $7,8$.

\begin{picture}(140,60)
\put(10,40){\line(1,0){15}}
\put(32.5,40){\oval(15,12)}
\put(40,40){\line(1,0){15}}
\put(25,40){\oval(30,25)[t]}
\put(40,40){\oval(30,25)[b]}
\put(10,40){\circle*{1}}
\put(25,40){\circle*{1}}
\put(40,40){\circle*{1}}
\put(55,40){\circle*{1}}
\put(32,15){\makebox(0,0){Fig. 7}}           %

\put (90,40){\oval(15,12)}
\put(105,40){\oval(15,12)}
\put(120,40){\oval(15,12)}
\put(82.5,40){\circle*{1}}
\put(97.5,40){\circle*{1}}
\put(112.5,40){\circle*{1}}
\put(127.5,40){\circle*{1}}
\put(105,15){\makebox(0,0){Fig. 8}}           %

\thicklines
\put(19,40){\vector(1,0){0}}
\put(49,40){\vector(1,0){0}}
\put(24,52.5){\vector(-1,0){0}}
\put(39,27.5){\vector(-1,0){0}}
\put(31,46){\vector(-1,0){0}}
\put(34,34){\vector(1,0){0}}

\put(88,34){\vector(-1,0){0}}
\put(91,46){\vector(1,0){0}}
\put(103,34){\vector(-1,0){0}}
\put(106,46){\vector(1,0){0}}
\put(118,34){\vector(-1,0){0}}
\put(121,46){\vector(1,0){0}}
\end{picture}

\vspace{-1cm}
Note that
the subgraph in $G$ induced by the set
$X\setminus\{z,y,u,v\}$ is complete symmetric, and each vertex in the
set $X\setminus\{z,y,u,v\}$ is connected to the set $\{z,y,u,v\}$ by
exactly six arcs.
Therefore, the vertex set of the graph $G$ can be
partitioned into subsets $\{z\}, X_1, X_2, X_3,X_4,\{v\}$ such that the
subgraph induced by the set $X_1\cup X_2\cup X_3\cup X_4$ is complete
symmetric, each vertex in the set $X_1$ is connected by a pair of
antiparallel arcs to the vertex $z$, each vertex in the set $X_2$ is
connected by a pair of antiparallel arcs to the vertex $v$, arcs go from
each vertex in the set $X_3$ to the vertices $z,v$, arcs go from the vertices
$z,v$ to each vertex in the set $X_4$; there are no other arcs in the
graph $G$. The decomposition is depicted in Figure 9.

\begin{figure}[hbpt]
\begin{picture}(130,60)
\put(65,40){\oval(30,30)[]}
\put(50,40){\line(1,0){30}}
\put(65,25){\line(0,1){30}}
\put(60,45){\makebox(0,0)[cc]{$X_1$}}
\put(70,45){\makebox(0,0)[cc]{$X_2$}}
\put(60,35){\makebox(0,0)[cc]{$X_3$}}
\put(70,35){\makebox(0,0)[cc]{$X_4$}}
\put(55,45){\line(-4,-1){20}}
\put(35,40){\line(4,-1){20}}
\put(75,45){\line(4,-1){20}}
\put(95,40){\line(-4,-1){20}}
\put(33,44){\makebox(0,0)[cc]{$z$}}
\put(97,44){\makebox(0,0)[cc]{$v$}}
\put(60,40){\oval(50,40)[lb]}
\put(58,30){\oval(20,20)[rb]}
\put(75,40){\oval(40,40)[rb]}
\put(77,30){\oval(30,20)[lb]}
\put(35,40){\circle*{1}}
\put(95,40){\circle*{1}}
\put(65,10){\makebox(0,0){Fig. 9}}

\thicklines
\put(50,20){\vector(-1,0){0}}
\put(82,20){\vector(1,0){0}}
\put(91,41){\vector(4,-1){0}}
\put(84,37){\vector(-4,-1){0}}
\put(82,43.3){\vector(-4,1){0}}
\put(44,37.7){\vector(-4,1){0}}
\put(39,41){\vector(-4,-1){0}}
\put(47.5,43.2){\vector(4,1){0}}
\end{picture}
\end{figure}

We note that some subsets in the decomposition can be empty, but the
relation $|X_1||X_2|+|X_3||X_4|>0$ must hold. Thus we have proved

\begin{Th} \label{th8}

For $k\ge 4$ maximal digraphs of quasi-diameter $k$ are either maximal
digraphs of radius $k$ or are obtained from them by reversing the
direction of all arcs.

For $k=3$ three cases arise.

1) Maximal digraph $D$ of quasi-diameter 3 is a maximal digraph of
radius 3;

2) Maximal digraph $D$ of quasi-diameter 3 is obtained from a maximal
digraph of radius 3 by reversing the direction of its arcs.

3) The vertex set of digraph $D$ can be partitioned into disjoint
subsets $\{z\}, X_1, X_2$, $X_3,X_4,\{v\}$ such that

\begin{minipage}{14cm}
\hspace{1cm}a) $|X_1||X_2|+|X_3||X_4|>0$;

\hspace{1cm}b) the subgraph induced by the set $X_1\cup X_2\cup
X_3\cup X_4$
is complete symmetric;

\hspace{1cm}c) each vertex in the set $X_1$ is connected by a pair of
antiparallel arcs to the vertex $z$, each vertex in the set $X_2$ is
connected by a pair of antiparallel arcs to the vertex $v$;

\hspace{1cm}d) arcs go from each vertex in the set $X_3$ to
each of the vertices $z,v$
and from each of the vertices $z,v$ to each vertex in the set $X_4$;
there are no other arcs in the graph $D$.
\end{minipage}
\end{Th}

Maximal digraph of quasi-diameter 2 is obtained from a complete
symmetric digraph by the removal of a pair of antiparallel arcs; and for
$d_m=1$ a complete symmetric digraph is the only maximal digraph.

Thus, we have obtained a description of all maximal digraphs of finite
quasi-diameter.

Let $\nu(n,k)$ denote the number of nonisomorphic maximal $n$-vertex
digraphs of quasi-diameter $k$, $\mu(n,k)$ denote the number of distinct
digraphs of quasi-diameter $k$ that can be constructed on $n$ numbered
vertices.

\begin{Cor} \label{cor61}
$$
\nu(n,k)=\left\{
\begin{array}{ll}
1 & \mbox{ for }1\le k\le 2,\\
\sfrac{(n-3)(n+4)}{2}+\dsum\limits_{t=1}^{n-4}\intt{\sfrac{t}{2}}
(n-t-1)+\intt{\sfrac{n-3}{2}} & \mbox{ for }k=3,\\
2(n-k-1)(k-2)+2 & \mbox{ for }k\ge 4.
\end{array}
\right.
$$
\end{Cor}

{\bf Proof.\/}
Theorem \ref{th6} and Corollary \ref{cor51} of Theorem \ref{th6} imply
that the assertion is true for $1\le k\le 2$ and $k\ge 4$. It is easy to see
that for $k=3$ the number of nonisomorphic nonbiconnected $d_m$-critical
digraphs equals $2n-6$. Hence it remains to count the number of
nonisomorphic biconnected $d_m$-critical graphs with $d_m=3$.
Let $G=(X,U)$ be a critical biconnected graph of quasi-diameter 3 and
$|X_1||X_2||X_3||X_4|>0$, $|X_1|+|X_2|=t$.

Considering Figure 9 we see that if we interchange the sets $X_1$ and
$X_2$, then we obtain a graph isomorphic to the original. Hence the
number of nonisomorphic biconnected critical graphs of
quasi-diameter 3, in which $|X_1||X_2||X_3||X_4|>0$ and $|X_1|+|X_2|=t$,
equals $\intt{\sfrac{t}{2}}(n-t-3)$, and the total number of such critical
digraphs in which $|X_1||X_2||X_3||X_4|>0$ equals
$\sum\limits_{t=2}^{n-4}\intt{\sfrac{t}{2}}(n-t-3)$. It is easy to see
that the number of nonisomorphic biconnected critical graphs with $d_m=3$
in which $|X_1||X_2|=0$ equals $\sum\limits_{t=2}^{n-2}(t-1)$, and the
number of such nonisomorphic graphs in which $|X_3||X_4|=0$ equals
$2\sum\limits_{t=2}^{n-3}\intt{\sfrac{t}{2}}+\intt{\sfrac{n-2}{2}}$.
Therefore,
\begin{eqnarray*}
\nu(n,3)&=&2n-6+\dsum\limits_{t=2}^{n-4}\intt{\frac{t}{2}}(n-t-3)+
 2\sum\limits_{t=2}^{n-3}\intt{\frac{t}{2}}+\intt{\frac{n-2}{2}}+
\sum\limits_{t=2}^{n-2}(t-1)=\\
&=& \frac{(n-3)(n+4)}{2}+\dsum\limits_{t=2}^{n-4}\intt{\sfrac{t}{2}}
(n-t-1)+\intt{\sfrac{n-3}{2}}. \ \ \rule{2mm}{2mm}
\end{eqnarray*}

\begin{Cor} \label{cor62}
$$
\mu(n,k)=\left\{
\begin{array}{ll}
1 & \mbox{ for }k=1,\\
\sfrac{n(n-1)}{2} & \mbox{ for }k=2,\\
n(n-1)(2^{2n-5}-2) & \mbox{ for }k=3,\\[3mm]
\hspace{-.7ex}
\begin{array}{l}
2\left\{(k-1)k!\binom{n}{k}+\right.\\
+\left.(k-2)(k-1)!\binom{n}{k-1}(2^{n-k+1}-2n+2k-4)\right\}
\end{array}
& \mbox{ for }k\ge 4.
\end{array}
\right.
$$
\end{Cor}

{\bf Proof.\/}
The assertion of the corollary for $1\le k\le 2$ and $k\ge 4$ is an
immediate consequence of Theorem \ref{th8} and Corollary \ref{cor52} of
Theorem \ref{th6}. It remains to prove the assertion for $k=3$.

By an
argument analogous to that used in the proof of Corollary \ref{cor61} we
can show that on given $n$ numbered vertices one can construct\\
$n(n-1)(2^{n-1}-4)$ nonbiconnected critical graphs of quasi-diameter 3;\\
$2n(n-1)(2^{n-3}-1)$ biconnected critical graphs with $d_m=3$\\
\phantom{$2n(n-1)(2^{n-3}-1)$ } in which $(|X_1|+|X_2|)(|X_3|+|X_4|)=0$;\\
$2\sum\limits_{t=2}^{n-3}n(n-1)\binom{n-2}{t}(2^t-2)$ biconnected
critical graphs of quasi-diameter 3 in which exactly one of the sets
$X_1,X_2,X_3,X_4$ is empty;\\
$\sum\limits_{t=2}^{n-4}n(n-1)\binom{n-2}{t}(2^{t-1}-1)(2^{n-t-2}-2)$
critical biconnected graphs in which $|X_1||X_2||X_3||X_4|>0$. \\
Here we used the same notation as in Figure 9. Thus we obtain that
\begin{eqnarray*}
&&\mu(n,k)=n(n-1)(2^{n-1}-4)+2n(n-1)(2^{n-3}-1)+\\
&&+2\sum\limits_{t=2}^{n-3}n(n-1)\binom{n-2}{t}(2^t-2)+
\sum\limits_{t=2}^{n-4}n(n-1)\binom{n-2}{t}(2^{t-1}-1)(2^{n-t-2}-2).
\end{eqnarray*}

One sees easily that the last expression equals $n(n-1)(2^{2n-5}-2)$.
\rule{2mm}{2mm}

\end{document}